\documentclass[manuscript]{acmart}
\acmSubmissionID{114}           

\usepackage{subcaption}
\usepackage[most]{tcolorbox}
\usepackage{tabularx,booktabs}

\acmConference[LAK '26]{Learning Analytics \& Knowledge}{Mar.~2026}{TBD}
\acmBooktitle{Proceedings of the 16th International Conference on Learning Analytics \& Knowledge (LAK '26)}
\acmYear{2026}
\acmDOI{}   
\acmISBN{}  

\title[When LLMs Fall Short in Deductive Coding]{When LLMs Fall Short in Deductive Coding: Model Comparisons and Human–AI Collaboration Workflow Design}

\author{Zijian Li}
\email{zijianli@stu.pku.edu.cn}
\affiliation{%
  \department{Department of Educational Technology}
  \institution{Graduate School of Education, Peking University}
  \city{Beijing}
  \country{China}
}

\author{Luzhen Tang}
\email{tangluzhen@pku.edu.cn}
\affiliation{%
  \department{Department of Educational Technology}
  \institution{Graduate School of Education, Peking University}
  \city{Beijing}
  \country{China}
}

\author{Mengyu Xia}
\email{marine\_xia@stu.pku.edu.cn}
\affiliation{%
  \department{Department of Educational Technology}
  \institution{Graduate School of Education, Peking University}
  \city{Beijing}
  \country{China}
}

\author{Xinyu Li}
\orcid{0000-0003-2681-4451}
\email{xinyu.li1@monash.edu}
\affiliation{%
  \department{Centre for Learning Analytics, Faculty of Information Technology}
  \institution{Monash University}
  \city{Melbourne}
  \country{Australia}
}

\author{Naping Chen}
\email{s\_chenpp@stu.edu.cn}
\affiliation{%
  \institution{Medical College, Shantou University}
  \city{Shantou}
  \country{China}
}

\author{Dragan Ga\v{s}evi\'{c}}
\orcid{0000-0001-9265-1908}
\email{dragan.gasevic@monash.edu}
\affiliation{%
  \department{Centre for Learning Analytics, Faculty of Information Technology}
  \institution{Monash University}
  \city{Melbourne}
  \country{Australia}
}

\author{Yizhou Fan}
\email{fyz@pku.edu.cn}
\authornote{Corresponding author.}
\affiliation{%
  \department{Department of Educational Technology}
  \institution{Graduate School of Education, Peking University}
  \city{Beijing}
  \country{China}
}

\begin{document}

\begin{abstract}
With generative artificial intelligence driving the growth of dialogic data in education, automated coding is a promising direction for learning analytics to improve efficiency. This surge highlights the need to understand the nuances of student-AI interactions, especially those rare yet crucial. However, automated coding may struggle to capture these rare codes due to imbalanced data, while human coding remains time-consuming and labour-intensive. 
The current study examined the potential of large language models (LLMs) to approximate or replace humans in deductive, theory-driven coding, while also exploring how human–AI collaboration might support such coding tasks at scale. We compared the coding performance of small transformer classifiers (e.g., BERT) and LLMs in two datasets, with particular attention to imbalanced head–tail distributions in dialogue codes. Our results showed that LLMs did not outperform BERT-based models and exhibited systematic errors and biases in deductive coding tasks. We designed and evaluated a human–AI collaborative workflow that improved coding efficiency while maintaining coding reliability. Our findings reveal both the limitations of LLMs -- especially their difficulties with semantic similarity and theoretical interpretations -- and the indispensable role of human judgment, while demonstrating the practical promise of human–AI collaborative workflows for coding.

\end{abstract}


\maketitle

\section{Introduction}

Coding of learning data represents a foundational step in learning analytics, as it enables researchers to systematically interpret learners' raw data and uncover underlying learning processes \cite{lang2022learning}. In learning analytics, raw data such as clickstreams, dialogue transcripts, and written assignments are often unstructured, making them challenging to analyse in original form. Coding provides the crucial bridge between raw data and theoretically meaningful constructs and allows researchers to capture learning processes, strategies, and cognitive or affective states in a systematic way \cite{neuendorf2017content}.

As the adoption of large language models in education grows, the scale of text-based interaction data has expanded rapidly \cite{yan2024promises}. Beyond numerical indicators, dialogic data (e.g., student–LLM conversations) require structured coding through codebooks to capture learning processes and states in a theoretically grounded manner \cite{cheng2025asking}. Yet qualitative coding can labour-intensive: multiple human coders must annotate, cross-check, and reconcile discrepancies to ensure reliability \cite{halpin2024inter}. With increasingly large and complex datasets, manual approaches are becoming infeasible and drive a growing interest in automated coding. Early work reframed coding as a supervised classification task, which applies rule-based classification, machine learning and deep learning models \cite{cai2019ncoder+,ferreira2020towards,wulff2023utilizing,gweon2024automated}. The emergence of transformer models, with BERT as a foundational example, significantly advanced automated coding \cite{iqbal2024towards}. This progression laid the groundwork for the current adoption of LLMs. However, these methods demand large, balanced training sets and often struggle to generalise across contexts and codebooks. Different codebooks and subtle differences in code definitions make transferring models across settings difficult without re‐annotation or adaptation \cite{dunivin2025scaling}.
These methods also suffer from cold-start problem -- i.e.,  when applying a new codebook, no labelled data are initially available to train the model, therefore requiring much more investment at the beginning \cite{than2025updating}.

Automatic coding methods face persistent challenges when applied to authentic educational datasets. One of the most prominent is the long-tail distribution of codes, where a small number of frequent codes dominate the data, while many rare but pedagogically important codes occur sparsely \cite{lin2023robust}. As a result, models trained on such data tend to overfit to high-frequency codes, achieving strong performance on the “head” (i.e., very frequent codes) of the code long-tail distribution while systematically neglecting the “tail” (i.e., infrequent) codes \cite{bujang2022imbalanced}. This imbalance is especially problematic but overlooked in learning analytics contexts \cite{sha2022leveraging}, where rare codes — such as indicators of metacognitive monitoring and critical reflection — often provide the most valuable insights into learners’ learning processes \cite{bistolfi2025towards}. Conventional imbalance-handling techniques (e.g., oversampling, SMOTE, cost-sensitive learning) improve predictive performance in general natural language processing tasks \cite{wongvorachan2023comparison,lin2023enhancing,sha2022leveraging}, but may fall short in detecting "tail" codes. The long-tail is not merely noise to be suppressed; it is a priority for preserving theoretically meaningful yet infrequent behaviours in learning analytics.


Recently, LLMs have attracted much attention for their potential to automate qualitative coding in education \cite{long2024evaluating} by demonstrating their semantic understanding capabilities and potentially capturing long-tail codes. Empirical studies have reported notable gains: GPT-4 has achieved high inter-coder agreement in classroom dialogue analysis and strong kappa values in interpretive tasks when prompted with reasoning \cite{long2024evaluating,dunivin2024scalable,dunivin2025scaling}. Hybrid approaches combining LLMs with codebooks have also shown the potential to save researchers' time and produce consistent and trustworthy coding codes \cite{liu2024assessing}. However, several limitations still persist, such as LLMs may introduce systematic biases, depend heavily on prompt design, and lack the interpretive depth needed for nuanced educational constructs \cite{ashwin2023using,chew2023llm,xiao2023supporting,oksanen2025llmcode,schroeder2025large,mizumoto2025large}. 

Inductive or thematic coding, where codes evolve iteratively from the data, has been the main area of application for LLMs. These coding tasks with generative capabilities of LLMs and have produced promising results in educational research \cite{schroeder2025large}.  Deductive coding, in contrast, requires consistent adherence to predefined, theory-driven codebooks, and here LLMs often struggle with semantic consistency and theoretical interpretation \cite{halterman2024codebook}. This limitation is most consequential in educational datasets, where the rare codes index theory-critical processes but occur rarely \cite{bistolfi2025towards,nye2015automated}. Deductive coding still remains comparatively underexplored, despite its importance for learning analytics, where rare but theoretically critical codes should be applied reliably at scale. Recent studies have started to investigated human–AI collaboration in qualitative analysis to address this issue and recent work on human–AI collaboration in qualitative analysis shows that LLMs can approximate human coding under certain conditions but remain constrained by patterned biases, prompt sensitivity, and limited interpretive depth \cite{chew2023llm,xiao2023supporting,ashwin2023using,oksanen2025llmcode}. Frameworks such as LACA, LLMCode and CollabCoder illustrate emerging approaches to integrate human engagement and LLMs, yet these systems are largely situated in inductive or thematic contexts. Far less is known about how such collaboration can be extended to deductive coding, where a predefined and theory-driven codebook must be applied consistently.


Collectively, studies on transformer architectures (e.g., BERT) and LLMs demonstrate clear potential for scaling qualitative coding, yet they also reveal unresolved concerns about reliability, validity, and applicability in deductive coding tasks. 
\textbf{In this study, we made three key contributions to advance deductive, codebook-guided coding in learning analytics:} (1) We conducted a systematic comparison between small transformer-based classifiers and LLMs under deductive, codebook-guided conditions, identifying both strengths and limitations. (2) We identified systematic error patterns in LLMs when used for deductive coding such as negativity bias and semantic confusions concentrated in codes, demonstrating why LLMs cannot simply replace human coders in deductive tasks. (3) We proposed a human–AI collaborative workflow that routes low-confidence and rare cases to human experts and evaluated to show improvements in the overall reliability and preservation of rare but pedagogically meaningful codes.  

\section{Background}

\subsection{Development of Automated Coding: From Rule-based Systems to LLMs}

Automated qualitative coding in learning analytics has progressed in several phases. Early systems, such as nCoder, used researcher-defined rules to scale coding, but they struggled to adapt when codebooks changed \cite{zambrano2023ncoder}. Furthermore, researchers began applying machine learning by framing coding as a text classification task. Machine learning and deep learning models, such as Random Forest (RF), Support Vector Machines (SVM), and Convolutional Neural Networks (CNN), have commonly been used for automated coding. Previous studies showed that linear classifiers and tree-based ensembles could achieve moderate accuracy with small labelled datasets, but deep learning models such as CNNs and recurrent networks provided notable improvements in accuracy of classification by capturing sequential and contextual information (e.g., first or last forum post in a thread) \cite{osakwe2022towards,ferreira2020towards, cavalcanti2019analysis,wang2017combining}, which may be needed during the process of coding. 

Transformer-based models further advanced automated coding. For instance, BERT has been adapted to various educational contexts, including classifying rhetorical codes in student essays \cite{iqbal2024towards} and teacher reflections \cite{wulff2023utilizing}, consistently outperforming earlier classifiers.
Yet such models typically depend on task-specific retraining or fine-tuning to handle new tasks. By contrast, contemporary LLMs leverage in-context learning to perform coding tasks directly from prompts, eliminating the need for retraining \cite{brown2020language}. This ability positions LLMs as practical solutions for coding tasks in learning analytics, where datasets are limited and codebooks are highly specialised. Unlike traditional supervised models, LLMs can be deployed in a cold-start manner, relying only on natural language definitions and examples.
For example, \citet{xiao2023supporting} demonstrated that GPT-3, when combined with expert-designed codebooks, achieved fair to substantial agreement with human coders in deductive curiosity-driven questions coding task.
\citet{liu2025qualitative} and \citet{than2025updating} evaluated GPT-4 across diverse educational datasets, confirming strong performance but noting variability across codes. \citet{long2024evaluating} compared GPT-4-annotated codes with human expert codes on classroom dialogues and found that GPT-4 substantially improved coding efficiency, achieving high inter-coder agreement, although systematic discrepancies or bias remained, particularly in coordination-related codes where the model relied on surface cues rather than broader context. Meanwhile, \citet{huang2025s} demonstrated that RoBERTa (a transformer-based model) outperformed GPT-4o in classifying peer feedback, underscoring the gap between domain-specific and general-purpose models. More recently, \citet{na2025llm} developed a multi-component framework to address the challenges of dialogic data. Their approach integrates structured prompting, ensemble reasoning and consistency checks, which substantially improved coding reliability in collaborative problem-solving transcripts \cite{na2025llm}. In addition, in a different disciplinary context, \citet{dunivin2024scalable} compared GPT-4 with GPT-3.5 in qualitative coding of historical news media passages and found that GPT-4 achieved substantially higher inter-coder reliability, whereas GPT-3.5 performed notably worse.

Overall, research shows a transition from rule-based systems to machine learning, transformer-based models, and LLMs in detective coding tasks in learning analytics. However, the findings are fragmented. Several studies report that smaller transformer-based models like BERT outperformed general-purpose LLMs \cite{huang2025s,zhang2025bert} or that their performance was largely comparable with no marked difference \cite{pilicita2025llms}. Others highlight GPT-4’s ability of natural language understanding \cite{dunivin2024scalable,long2024evaluating,liu2025qualitative}. However, few studies offer systematic comparisons across BERT and LLMs and different coding tasks, making it difficult to evaluate trade-offs and guide model selection for deductive coding in learning analytics.

\subsection{Long-tail and Imbalanced Data Distribution in Learning Analytics}
Automated coding in learning analytics faces a challenge: the imbalanced, long-tailed distribution of codes \cite{bistolfi2025towards,lin2023enhancing,sha2022leveraging}. In many datasets, a few frequent codes dominate, while rare but pedagogically important codes may appear scarcely. This skew may hinder automation, as the “tail” codes can carry high theoretical value but can be hard for models to find.
Results of \citet{bistolfi2025towards} showed that rare codes such as metacognitive monitoring and elaboration were so rare in student essays leading popular machine learning models to achieve low Cohen's $\kappa$ of about 0.2. Similarly, \citet{nye2015automated} annotated 1,438 tutoring sessions with 126 dialogue acts and 16 modes. While frequent codes were automatically coded at a human-level reliability, complex and rare codes performed much worse \cite{nye2015automated}. Comparable findings arise in research on automatic coding of discussions on MOOC forums, where urgent posts or instructor-intervention signals occur infrequently but matter most for timely support \cite{alrajhi2024solving}. In each case, skew biases classifiers toward majority codes, producing models that look accurate overall yet fail on the codes of critical value \cite{lin2023robust,sha2022leveraging}. 

To mitigate the limitations in automatic coding of rare codes, researchers have proposed a range of imbalance-handling strategies. Data-level methods such as oversampling and synthetic data generation attempt to balance distributions; algorithm-level methods such as cost-sensitive learning and reweighted loss functions adjust the training objective; hybrid ensemble approaches combine both \cite{bujang2022imbalanced,chen2024survey}. These approaches work well in general text classification, but results in educational coding are mixed. \citet{alrajhi2024solving} applied undersampling and data augmentation to improve the accuracy of classification in rare codes but found a significant drop in frequent codes. Rare codes often capture complex, context-dependent constructs that are hard to augment or weight effectively. Recent work points to robustness-oriented objectives. For example, \citet{lin2023robust} showed that optimising dialogue act classifiers with AUC-maximisation significantly improved overall performance in low-resource, imbalanced settings, underscoring the role of loss function design, but did not report the per-code performance. \citet{pratama2021imbalanced} used SMOTE-series resampling methods to improve the performance of ML models on an imbalanced educational dataset, but also did not report the detailed performance of rare codes. Effectively handling long-tail code distributions in authentic educational contexts remains an open challenge. Unlike other NLP domains, the tail in educational datasets is not noise to be smoothed out but the analytic priority, as it may index rare yet theoretically meaningful processes\citet{bistolfi2025towards}. Addressing this challenge may require a hybrid workflow that combines statistical robustness with human interpretive oversight, ensuring that pedagogically critical codes are preserved rather than marginalised.

\subsection{Human-AI Collaboration in Qualitative Text Coding: Promise and Limitations}

Recent studies demonstrate both the promise and the limits of LLMs-assisted coding. LLMs can achieve fair-to-substantial agreement with human coders in deductive tasks, particularly when supported by prompt design or reasoning chains \cite{chew2023llm,xiao2023supporting,long2024evaluating}. However, scholars have also highlighted broader limitations of LLMs-assisted coding. \citet{ashwin2023using} warned that LLMs introduce patterned biases that threaten validity and reliability. \citet{chew2023llm} proposed the LLMs-Assisted Content Analysis (LACA) framework, showing that while GPT-3.5 could achieve fair agreement with human coders in deductive coding tasks, it also tended to guess in ways that obscured when predictions were unreliable. Similarly, \citet{xiao2023supporting} demonstrated that GPT-3 combined with expert-developed codebooks reach fair-to-substantial inter-coder agreement, but noted that task simplification and prompt design critically limited performance. \citet{oksanen2025llmcode} introduced LLMCode to evaluate and enhance researcher–AI alignment, concluding that LLMs worked reasonably well in deductive coding but still lacked deep interpretive capacity. From a broader perspective, \citet{schroeder2025large} emphasized that the use of LLMs in qualitative research raises unresolved ethical questions and risks eroding researcher values such as deep engagement with data, commitment to interpretive multiplicity, and safeguarding of privacy, while \citet{mizumoto2025large} showed that LLMs could fall short in classifying learners’ open-ended responses, underscoring limits the use of LLMs for automated coding in education. Collectively, these studies indicate that while LLMs scale qualitative coding, challenges of reliability, validity, and interpretive depth remain unresolved. 

The limitations of LLMs in qualitative coding call for the value of human experts. Particularly, deductive coding is not a simple text–label match; it applies learning theory to situated language. Codebooks operationalise codes but remain incomplete instruments for representing them \cite{ritchie2022development}. Theory construction permeates the entire process of deductive qualitative research, as researchers engage in theory building both when operationalising guiding theories and when analysing data \cite{fife2024deductive}. In practice, experienced coders rely on theory-informed, partly tacit criteria to adjudicate boundary cases; for instance, considering missing but expected cues, integrating cross-turn coherence, or privileging inferential aims. Such judgment and coding strategy cannot be fully captured by prompts or rules, and those may be the key cues for detecting rare codes. \citet{zhang2023qualigpt} also emphasises that researchers or users must always retain control over the accuracy of results, with ultimate decision-making authority remaining centred on humans. \citet{sabbaghan2024exploring} also argues that the depth of interpretation and contextual understanding provided by human researchers remains irreplaceable.

To address gaps of LLMs and integrate the ability of human experts, studies have proposed systems that integrate iterative human–machine workflows. PaTAT \cite{gebreegziabher2023patat} allows researchers to split, merge, and refine codes in time, with models updating predictions accordingly. LLMCode \cite{oksanen2025llmcode} supports bi-directional alignment through iterative prompt and example refinement. TAMA \cite{xu2025tama} employs multi-agent ensembles with human feedback to synthesise thematic codes. CollabCoder \cite{gao2024collabcoder} facilitates collaboration across open coding, consensus building, and codebook creation, ensuring transparency. Finally, the LLM-in-the-loop approach \cite{dai2023llm} combines few-shot examples with iterative human revisions to construct evolving codebooks. However, most of these approaches have been developed in the context of inductive or thematic analysis, where codebooks were built and refined iteratively. By contrast, deductive coding, where a predefined, theory-driven codebook must be applied consistently, has received far less attention. Addressing this challenge requires workflows combining the efficiency of automated models, the semantic adaptability of LLMs, and the interpretive rigour of human expertise. Building on these observations, this study addressed the following two research questions:  

\begin{itemize} 
    \item \textbf{RQ1:} How do different transformer models perform deductive, codebook-guided coding across both frequent and rare codes? 
    \item \textbf{RQ2:} To what extent can human-AI collaborative workflows perform deductive coding reliably?   
\end{itemize}

\section{Methods}

\begin{figure}[htbp]
  \centering
  \includegraphics[width=\linewidth]{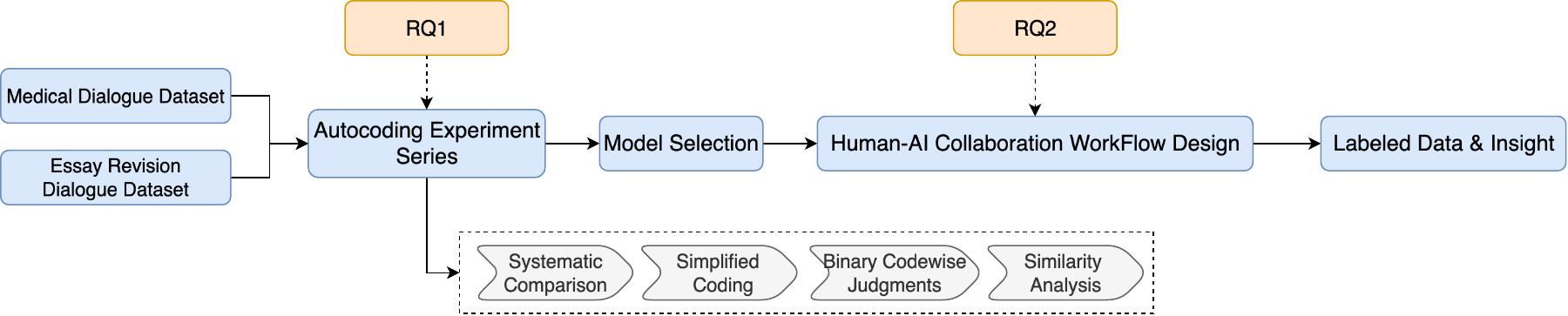}
  \caption{Overall Research Design}
  \Description{overall research design}
  \label{fig:overall}
\end{figure}

The overall research design is illustrated in Figure~\ref{fig:overall}. To address the research questions, this study employed two datasets: the medical dialogue dataset served as the primary source of analysis, while the essay revision dialogue dataset was introduced to further examine LLMs coding capabilities and to investigate potential impact of coding similarity on model performance. A series of automated coding experiments was first conducted to answer Research Question 1, focusing on performance differences across models and exploring the depth of LLMs’ coding capabilities. The findings from these experiments also informed the choice of models for Research Question 2, which focused on the value of human–AI workflows. Finally, we developed a workflow for human–AI collaboration that aims to ensure both efficiency and accuracy in automated coding. This workflow was empirically evaluated through a human study. 

\subsection{Dataset}
The medical dialogue dataset (in Chinese) was collected in 2024 through an experimental study conducted by previous research in a Medical School. Over a period of five weeks, medical students engaged in weekly Virtual Standardised Patient (VSP) history-taking tasks using a online Moodle-based platform that researchers developed, where ChatGPT was configured to simulate a VSP using a tailored prompt. Each week, students were presented with a new clinical case scenario. A total of 210 participants were enrolled in the study. Across the five-week period, a total of 77,420 valid dialogue entries were recorded and retained for analysis, with each week ranging from 14,000 to 18,000 entries.

\textbf{Codebook 1: History Taking.}  
Following data collection, the dialogue data underwent a cleaning process before being subjected to qualitative coding. Two medical experts annotated the data using the structured coding scheme of history taking developed by (AUTHOR BLINDED FOR REVIEW). This scheme, constructed via systematic text condensation, comprises 12 codes spanning three dimensions — clinical reasoning, information gathering, and communication (see Table~\ref{tab:codebook_history}). It has demonstrated strong inter-coder reliability ($\kappa = 0.95$–$1.00$) and construct validity in capturing students’ reasoning and information-gathering behaviours during history-taking with VSPs. For clarity, we provide an abbreviated version of the codebook in Table~\ref{tab:codebook_history}, while the complete definitions and illustrative examples are included in Appendix~\ref{appendix:codebook_history} in \href{https://drive.google.com/drive/folders/1e0Qd_Lz0XGPmd4tpY7Inkvq4sV4zQA1M?usp=drive_link}{\textcolor{blue}{the Supplementary Materials (clickable link)}}.

The distribution of medical dialogue codes exhibited a typical long-tailed imbalance. A few codes (e.g., SS, LO, RQ, CC) dominated the dataset, while many codes occurred only rarely. The distribution was dominated by RQ (54.2\%) and SS (21.8\%), followed by LO (8.5\%) and CC (8.1\%), with all remaining codes each constituting less than 3\%. 
Accordingly, we adopt 5\% as the cut-off between “head” and “tail” codes in this study. Note that this threshold was driven by the empirical distribution in our data rather than by a prior canonical standard in the learning analytics literature, and this choice should be regarded as a heuristic.
We identify the following codes as rare labels: \textit{PQ}, \textit{OS}, \textit{RR}, \textit{CK}, \textit{FQ}, \textit{SI}, \textit{RT}, and \textit{SR}. These rare codes (e.g., PQ, RR, SI) are of particular analytic importance, as they often capture subtle reasoning or information-gathering processes (Author, 2025). In addition to the medical dialogue dataset, we also used another codebook and the dataset used in a prior published study, which is in English \citet{cheng2025asking}. This codebook and dataset allowed us to test whether the results generalise across different codebooks and datasets.

\begin{table*}[htbp]
  \centering
  \small
  \caption{Codebook 1: History Taking (abbreviated with examples, adapted from (Author, 2025)).}
  \label{tab:codebook_history}
    \resizebox{0.9\textwidth}{!}{
  \begin{tabular}{lll}
    \toprule
    \textbf{Code} & \textbf{Short Definition} & \textbf{Example} \\
    \midrule
    PQ & Pathophysiologic reasoning question & ``Does the pain get worse when climbing stairs?'' \\
    RR & Relevant response to diagnostic info & ``For how long have you been taking contraceptives?'' \\
    SI & Summarizing and integrating information & ``You had sudden right chest pain this morning after blowing a balloon...'' \\
    LO & Logically organized diagnostic question & ``Do you have shortness of breath or cyanosis?'' \\
    SS & Specifying symptom characteristics & ``What triggered your chest pain today?'' \\
    RQ & Routine background question & ``Do you have any other discomfort?'' \\
    SR & Simple summarizing or restating & ``You came to the hospital for chest pain and shortness of breath...'' \\
    CK & Checking or confirming patient info & ``So the pain lasted about 2 minutes, right?'' \\
    RT & Repeating previously asked question & ``Do you have shortness of breath?'' (already asked earlier) \\
    FQ & Vague or fuzzy question & ``Any other diseases?'' \\
    CC & Chitchat or reassurance & ``Hello, I’m Dr. Zhang. Don’t worry, take your time.'' \\
    OS & Off-topic or irrelevant statement & ``Did the balloon get sucked in?'' \\
    \bottomrule
  \end{tabular}}
\end{table*}


\textbf{Codebook 2: Question Types and Mechanisms.}  
The coding scheme of the second dataset classified student questions into 14 codes, distinguishing between shallow-level inquiries (e.g., verification, definition), deep-level reasoning (e.g., causal consequence, instrumental), and requests (direct or indirect). Codes, definitions, and representative examples are provided in Appendix~\ref{appendix:codebook_combined} in \href{https://drive.google.com/drive/folders/1e0Qd_Lz0XGPmd4tpY7Inkvq4sV4zQA1M?usp=drive_link}{\textcolor{blue}{the Supplementary Materials}}. In addition, the scheme also classifies question mechanisms by their pragmatic mechanism, focusing on knowledge, coordination, and conversational management.

This supplementary codebook differs from the medical dialogue coding scheme in two respects: (i) Codes of mechanism were lexically distinct and semantically separable, and (ii) the codebook has been validated in prior learning analytics research \cite{cheng2025asking}. 
Prior work suggests that LLMs show variable performance on deductive coding tasks \cite{liu2025qualitative,long2024evaluating,huang2025s}. By testing LLMs on these datasets, we examined whether such challenges arise primarily from emantic overlap among codes, or from more fundamental difficulties in adhering to deductive coding schemes.

\subsection{Model Selection and Evaluation}

This study employed four small-scale transformer-based models and five large-scale language models to evaluate their performance in coding history-taking dialogue data. We fine-tuned four transformer-based models using standard hyperparameters with class reweighting and frequency-based sampling to mitigate long-tail imbalance. These models are commonly used baselines in Chinese NLP benchmarking (e.g. RoBERTa-wwm-ext in Chinese text classification tasks, MacBERT in Chinese pretraining studies) and are known to fine-tune effectively under moderate resource conditions \cite{xu2021roberta}. Five LLMs (Qwen3-8B, Qwen3-30B, DeepSeek-V3.1, GPT-4o-mini, GPT-4o) were evaluated in zero-shot settings through inference APIs, reflecting typical cold-start deployment in educational contexts.

\subsection{RQ1: Comparing Model Performance and Probing LLM Capability}\label{sec:exp_rq1}

\textbf{Experiment 1: Model Comparison.}  
We compared transformer classifiers and LLMs on the week-2 dataset, focusing on frequent vs. rare codes.\textbf{Experiments 2–4: Probing LLMs.}  
To test reliance on codebook definitions and semantic overlap, we:  
(1) simplified the code scope for 1,000 rare-code turns to reduce prompt length and ambiguity ;  
(2) reframed coding as binary judgments on 500 turns (6,000 decisions in total) to tested whether LLMs truly relied on codebook definitions;  
(3) analysed embeddings (100 samples per code, 12 codes) via PCA visualisation to examine the semantic similarity per code.  
All were evaluated with Cohen’s $\kappa$ or confusion matrices. \textbf{Supplementary Tests.}  
A second dataset confirmed generalizability using full codebooks and embedding similarity to test generalizability beyond single dataset.  
Prompt templates are in Appendix~\ref{sec:appendix_prompts} in \href{https://drive.google.com/drive/folders/1e0Qd_Lz0XGPmd4tpY7Inkvq4sV4zQA1M?usp=drive_link}{\textcolor{blue}{the Supplementary Materials}}.

\begin{figure}[htbp]
  \centering
  \includegraphics[width=0.9\linewidth]{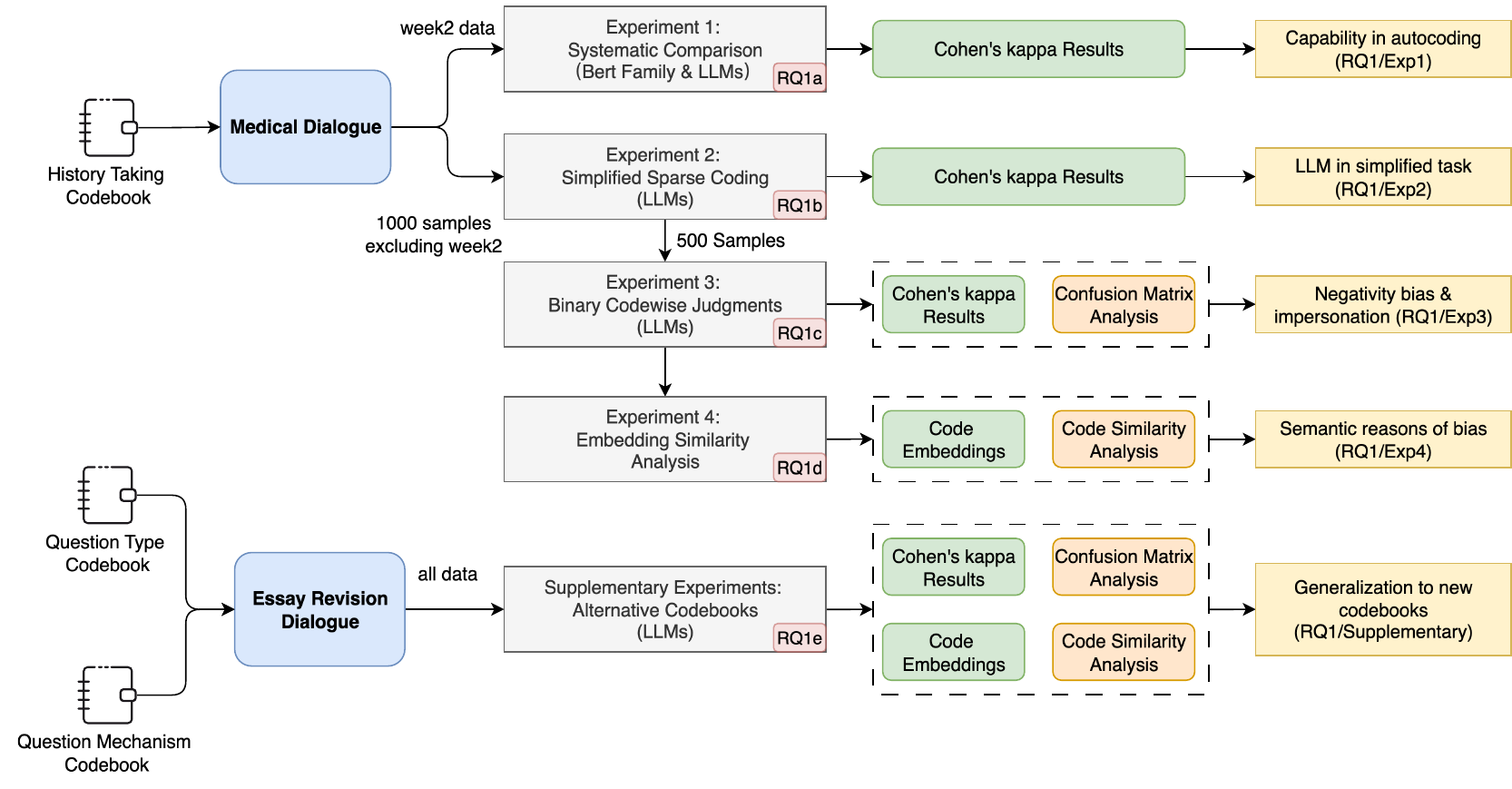}
  \caption{Overall experiment design and analyses (RQ1).}
  \Description{experiment design}
  \label{fig:exp_design}
\end{figure}

\subsection{RQ2: Human–AI Collaboration Workflow}

To answer RQ2, we designed an experiment to evaluate the effectiveness of a hybrid human-AI workflow in deductive coding. The goal was to examine whether combining transformer-based models, LLMs, and human experts can improve overall coding reliability and ensure the preservation of rare but theory-critical codes.

\begin{figure}[htbp]
  \centering
  \includegraphics[width=0.8\linewidth]{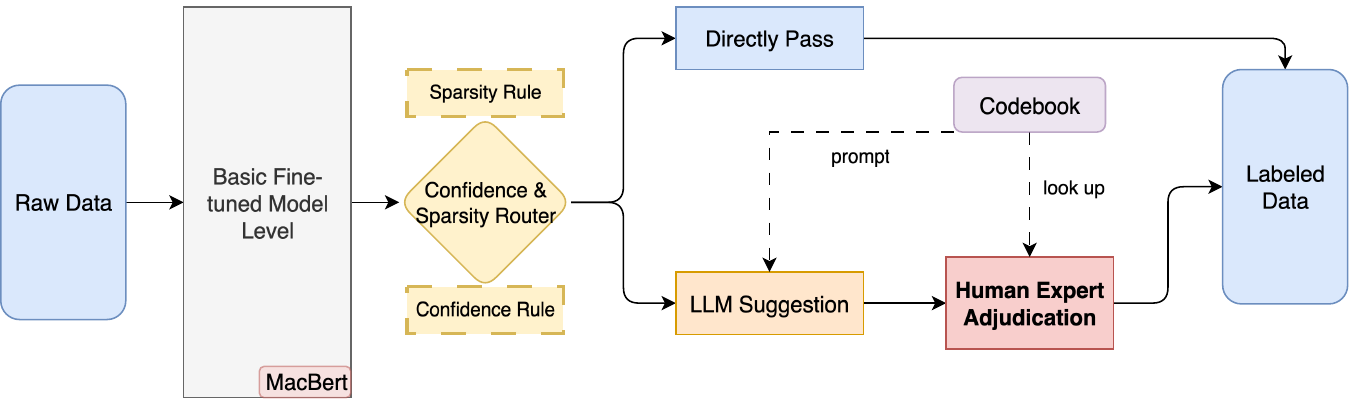}
  \caption{Human-AI Collaboration Workflow Design for RQ2}
  \Description{experiment design}
  \label{fig:workflow}
\end{figure}

The workflow consisted of three sequential stages (Figure~\ref{fig:workflow}):  
1. \textbf{Automated coding by MacBERT.} A fine-tuned MacBERT model from RQ1 and produced initial codes for all samples.  
2. \textbf{Confidence and sparsity routing.} Codes were routed based on two conditions: (i) low confidence (probability $<0.6$), or (ii) assignment to a rare code category (occurring in $<5\%$ of samples).  
3. \textbf{LLMs-assisted expert adjudication.} Escalated cases were passed to GPT-4o, which generated candidate codes and rationales. These suggestions were not accepted automatically; rather, they were presented alongside the MacBERT output, codebook definitions, and dialogue context for human expert adjudication. The base model used in the workflow was the MacBERT fine-tuned on the week-2 dataset of medical history-taking dataset, which demonstrated the most balanced performance across codes. For evaluation, we randomly sampled 500 dialogue turns from the week-3 dataset. After passing through the confidence-and-sparsity router, 44 samples were flagged for escalation and expert adjudication. The expert correction of these 44 cases required approximately 45 minutes in total, and reliability was assessed using Cohen’s $\kappa$ between the final results of workflow and ground-truth.

\section{Results}
\subsection{RQ1: How do small transformer classifiers and LLMs differ on deductive, codebook-guided coding across head vs. tail codes?}

\paragraph{RQ1a. Systematic model comparison.}

Table~\ref{tab:model_comparison} summarises Cohen’s $\kappa$ for three transformer baselines and four LLMs across twelve codes. Small-scale transformers (BERT, RoBERTa, DistilBERT) consistently outperformed LLMs on frequent codes (e.g., SS, LO, RQ, $\kappa > 0.85$). LLMs showed weaker and more variable performance overall, though larger-parameter models (GPT-4o, DeepSeek-V3.1) surpassed smaller ones (Qwen3-8B, GPT-4o-mini). Notably, LLMs displayed relative strengths in certain mid-frequency codes (e.g., SI, CC), suggesting that semantic adaptability may be beneficial for codes whose definitions are more context-dependent in conversational data. However, all models struggled on rare codes, where $\kappa$ often approached zero. Taken together, BERT-based models remain more reliable for deductive coding under structured schemes, while LLMs offer only partial strengths. In the optimisation step, we refined the transformer baselines by replacing DistilBERT with MacBERT and adapting RoBERTa to a Chinese variant. With loss reweighing and weighted sampling, these models achieved clear gains on long-tail codes: $\kappa$ rose from near zero to 0.18–0.36 on CK, SR, and OS, and exceeded 0.60 on moderately rare codes such as PQ and FQ. Frequent codes remained stable at $\kappa > 0.83$, confirming that optimisation improved rare-code reliability without sacrificing head performance (Appendix~\ref{appendix:bert_optimized} in \href{https://drive.google.com/drive/folders/1e0Qd_Lz0XGPmd4tpY7Inkvq4sV4zQA1M?usp=drive_link}{\textcolor{blue}{the Supplementary Materials}}).

\begin{table*}[htbp]
  \centering
  \caption{Comparison of Cohen’s $\kappa$ between transformer-based models and LLMs across codes.}
  \label{tab:model_comparison}
  \begin{tabular}{lcccccccc}
    \toprule
    \textbf{Code} & \textbf{BERT} & \textbf{DistilBERT} & \textbf{RoBERTa} & \textbf{Qwen3-8B} & \textbf{Qwen3-30B} & \textbf{DeepSeek-V3.1} & \textbf{GPT-4o} \\
    \midrule
    SS   & 0.90 & 0.89 & 0.86 & 0.40 & 0.54 & 0.37 & 0.55 \\
    LO   & 0.84 & 0.84 & 0.78 & 0.24 & 0.36 & 0.56 & 0.63 \\
    RQ   & 0.90 & 0.89 & 0.85 & 0.57 & 0.71 & 0.66 & 0.75 \\
    PQ   & 0.56 & 0.55 & 0.00 & 0.11 & 0.19 & 0.15 & 0.15 \\
    RR & 0.51 & 0.57 & 0.00 & 0.12 & 0.19 & 0.13 & 0.35 \\
    SI  & 0.79 & 0.80 & 0.74 & 0.62 & 0.45 & 0.76 & 0.50 \\
    SR  & 0.00 & 0.00 & 0.00 & 0.00 & 0.07 & 0.00 & 0.19 \\
    CK   & 0.00 & 0.06 & 0.00 & 0.17 & 0.16 & 0.13 & 0.25 \\
    FQ   & 0.26 & 0.41 & 0.00 & 0.00 & 0.16 & 0.14 & 0.11 \\
    RT   & 0.00 & 0.00 & 0.00 & 0.00 & 0.00 & 0.09 & 0.13 \\
    CC   & 0.93 & 0.93 & 0.88 & 0.58 & 0.65 & 0.73 & 0.75 \\
    OS   & 0.00 & 0.00 & 0.00 & 0.00 & 0.10 & 0.13 & 0.11 \\
    \bottomrule
  \end{tabular}
\end{table*}

\paragraph{RQ1b. Capability probing with LLMs.}

To further explore the coding ability of LLMs, we reduced the coding scope. Appendix Table~\ref{tab:exp1_llm} shows Cohen’s $\kappa$ for GPT-4o and GPT-4o-mini when the task was restricted to rare codes only, with prompts providing definitions and one positive example for those codes. Performance remained modest overall, with GPT-4o slightly outperforming GPT-4o-mini on most codes. The strongest agreement was observed for CC ($\kappa=0.71$) and FQ ($\kappa=0.56$), while codes such as RT and OS remained weak ($\kappa < 0.20$). GPT-4o-mini achieved higher agreement than GPT-4o on SI ($\kappa=0.62$ vs. 0.29), suggesting variability in code-level alignment even under reduced coding scope. These results indicate that narrowing task alleviated but did not eliminate LLM difficulties with rare codes.

Based on the low reliability in experiment~1, we conducted a more simplified task of making binary judgments per code. Appendix Table~\ref{tab:exp2_llm} reports results when each dialogue turn was evaluated against all 12 codes as binary yes/no judgments. Compared to Experiment~1, performance dropped sharply, with most rare codes below $\kappa=0.20$. Only SI ($\kappa \approx 0.50$) and CC ($\kappa \approx 0.55$) retained moderate reliability. Reasoning-related codes (PQ, RR) were identified with weak reliability ($\kappa < 0.45$). For RQ, SS, and LO, $\kappa=0$ reflected the absence of positives in the ground truth rather than model failure. Confusion matrices (Figure~\ref{fig:exp2_confusion}) show that both GPT-4o and GPT-4o-mini exhibited a potential negativity bias: about 60\% of outputs were “no,” reflecting the low base rate of true codes per turn. Errors were systematic rather than random. False positives concentrated on a few codes such RQ, LO, SS (frequent codes) and RR, CK (rare codes), with consistent patterns across both models (Figure~\ref{fig:exp2_fp_error_distribution_in_code} in the appendix). To further analyse these errors, we examined the true labels behind false positives (Appendix~\ref{appendix:impersonation}). This substitution analysis revealed that PQ was most frequently mislabeled as RQ, LO, or SS, with some spillover into RR and CK. These results indicate a systematic confusion pattern surrounding PQ: models tend to overassign broader or semantically related codes, thereby obscuring PQ’s more fine-grained meaning.

\paragraph{RQ1c. Embedding similarity analyses.}


To further investigate whether systematic misclassifications in Experiment~2 could be explained by inherent semantic overlap among codes, we computed sentence embeddings for 50 representative student questions per code and calculated pairwise cosine similarities across all 12 codes. The results revealed a high degree of semantic closeness, with an average similarity of $0.857$ (\textit{SD} $=0.0736$). This indicates that most codes were lexically and semantically close to one another, making reliable discrimination difficult and suggesting that the boundaries between codes are not sharply defined.

\begin{figure}[htbp]
  \centering
  \begin{subfigure}{0.45\linewidth}
    \includegraphics[width=\linewidth]{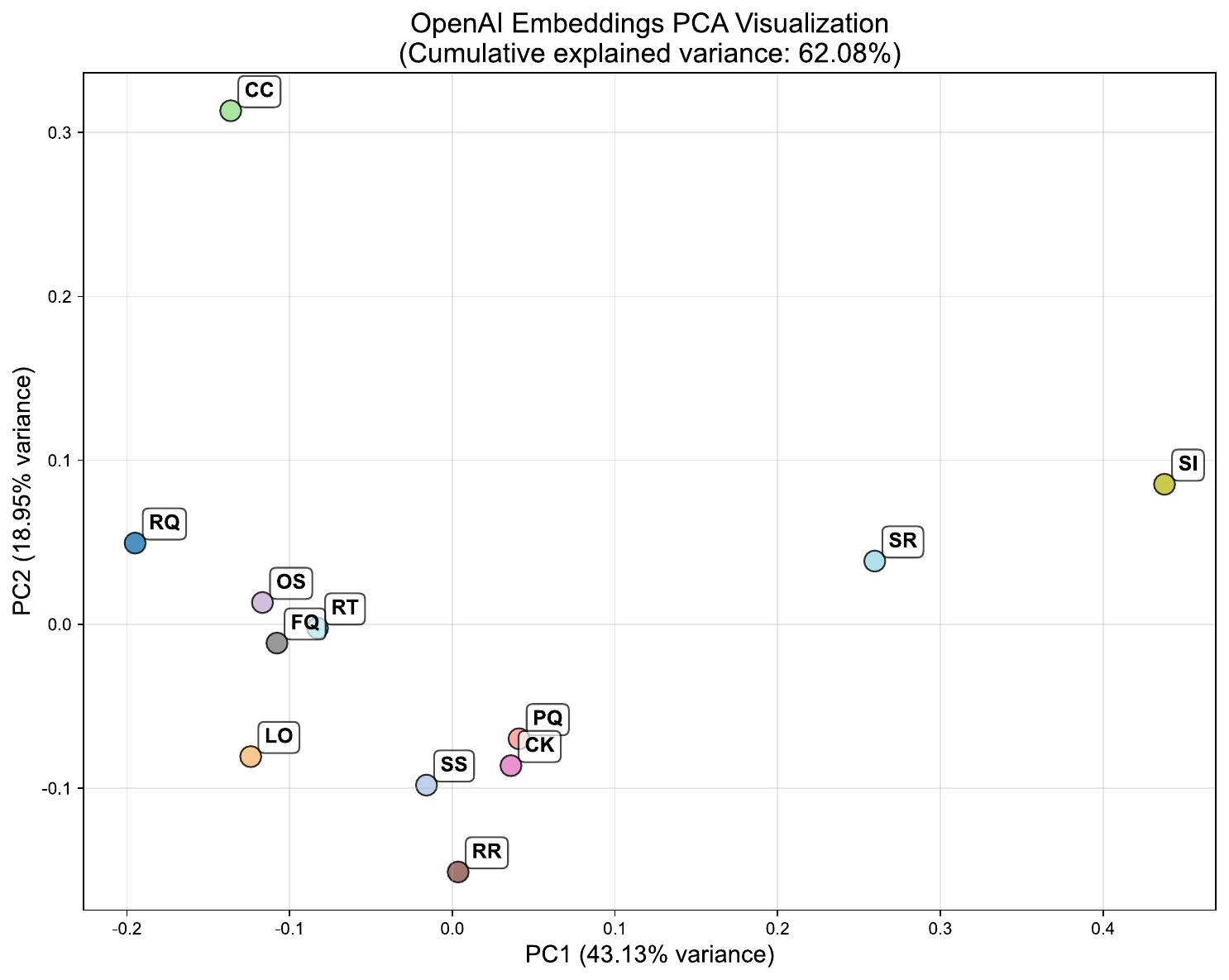}
    \caption{PCA Scatter of Codes}
    \label{fig:exp3_pca_scatter}
  \end{subfigure}
  \hfill
  \begin{subfigure}{0.45\linewidth}
    \includegraphics[width=\linewidth]{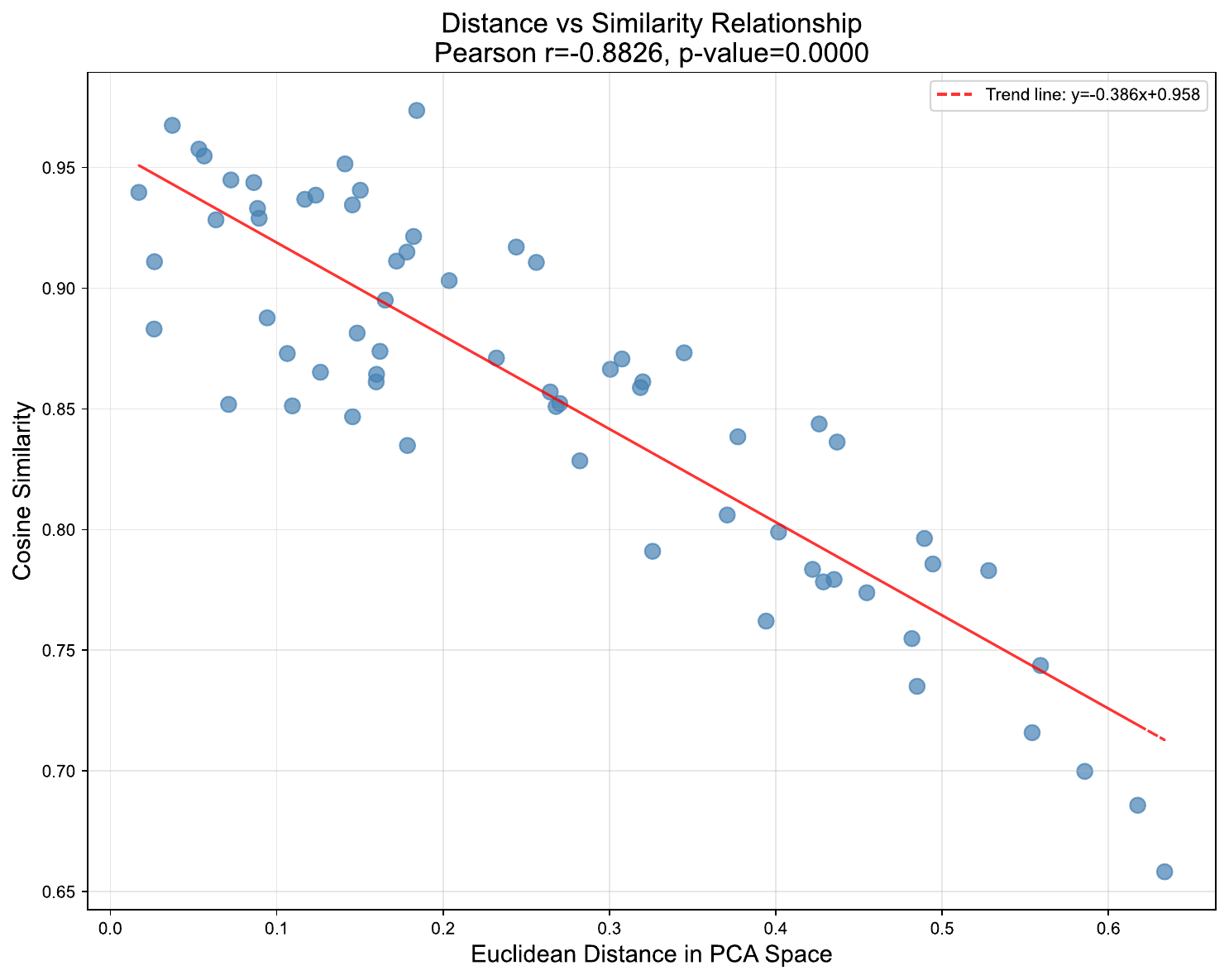}
    \caption{Similarity-Distance}
    \label{fig:exp3_similarity_distance}
  \end{subfigure}
  \caption{Embedding Similarity Analysis Result.}
  \label{fig:exp3_embedding_similarity}
\end{figure}

To visualise these relationships, we applied PCA to reduce the original 3,072-dimensional embeddings into two principal components. As shown in Figure~\ref{fig:exp3_pca_scatter}, the first component explained 43.13\% of the variance and the second explained 18.95\%. The scatter plot (panel a) highlights clusters of semantically adjacent codes. For example, \textbf{PQ} is positioned closest to \textbf{SS}, \textbf{CK}, and \textbf{RR}, and is not far from \textbf{LO}. This spatial proximity aligns with the impersonation errors observed in Experiment~2, where PQ was frequently misclassified as one of these codes. In addition, the relationship between embedding distance and cosine similarity was strongly negative, as expected: Pearson’s $r = -0.8826$, $p < .001$ (Figure~\ref{fig:exp3_similarity_distance}). This confirms that the PCA distances provide a meaningful representation of inter-code semantic similarity.

\paragraph{RQ1d. Supplementary codebooks (question types \& mechanisms).}
To test generalizability beyond the medical dialogue dataset, we used a supplementary codebook with the second dataset to examine whether the results hold across different coding contexts. Appendix Tables~\ref{tab:supp_question_types} and~\ref{tab:supp_mechanisms} summarise results on the supplementary codebook. For question types, codes with clear operational boundaries achieved higher agreement, with DeepSeek-V3.1 and GPT-4o often leading. In contrast, several low-frequency types (e.g., \emph{Off Topic}, \emph{Expectational}) yielded $\kappa \approx 0$, underscoring the instability of LLM judgments under data scarcity. A notable case was \emph{Judgmental}, the most frequent type in this dataset (support = 163), yet still poorly recognised ($\kappa \leq 0.44$). Confusion matrices (Appendix Figure~\ref{fig:fn_judgemental}) show systematic misclassifications with \emph{Direct Request}, \emph{Indirect Request}, and \emph{Verification}. Embedding similarity analysis (Figures~\ref{fig:qt_embedding_similarity}–\ref{fig:mechanism_embedding_similarity}) showed that \emph{Judgmental}, \emph{Direct Request}, \emph{Indirect Request}, and \emph{Verification} cluster closely, indicating that their semantic overlap likely contributes to the observed misclassification. For question mechanisms, performance was moderate-to-substantial for well-supported codes (\emph{Knowledge Deficit}: $\kappa=0.51$–$0.73$; \emph{Social Coordination}: $\kappa=0.39$–$0.59$). Embedding visualisations show that the four mechanisms are semantically well-separated, yet model agreement remains modest.

Overall, even with semantically distinct alternative codebooks, LLMs failed to deliver consistent, codebook-adherent decisions. This suggests that the difficulty lies not only in semantic confusability but also in the models’ limited ability to operationalise and apply deductive, theory-bound criteria, particularly for rare codes.

\subsection{RQ2: How can human–AI collaboration support scalable deductive coding? }

\paragraph{Overall workflow performance.}  
At the aggregate level, the baseline MacBERT model achieved $\kappa=0.62$. 
Replacing human judgment with LLMs-only predictions, reliability remained at $\kappa=0.62$, confirming that current LLMs cannot serve as a drop-in substitute for expert coding. However, incorporating expert adjudication into the workflow increased reliability to $\kappa=0.66$, with 21 corrections made. In an idealised scenario where all low-confidence cases were reviewed, reliability could reach $\kappa=0.70$, representing a 0.08 improvement over the baseline.


\paragraph{Code-level improvements.}  
Table~\ref{tab:rq2_codewise} reports the impact of the workflow on each medical history-taking code. The largest improvement occurred for the rare but theory-critical code \textit{FQ}, where reliability increased by 0.20 after expert adjudication. Moderate gains were also observed for \textit{RQ} (+0.05) and \textit{PQ} (+0.06), both reflecting reasoning and information-gathering processes central to clinical competence. In contrast, high-frequency surface-level codes, such as \textit{SS}, improved only marginally (+0.01), and extremely rare codes (\textit{RT}, \textit{RR}, \textit{SI}, \textit{SR}) showed no measurable gain due to the insufficient number of positive samples. Interestingly, \textit{OS} exhibited a slight positive adjustment (+0.01), though the overall agreement remained very low. These results highlight that the workflow is especially valuable for preserving and refining low-frequency, yet theoretically meaningful codes, which are often neglected in automated pipelines.

\begin{table}[htbp]
  \centering
  \caption{Cohen’s $\kappa$ before and after applying the human--AI workflow for each medical history-taking code.}
  \label{tab:rq2_codewise}
    \resizebox{0.9\textwidth}{!}{
  \begin{tabular}{lcccccc}
    \toprule
    \textbf{Code} & \textbf{\# True Samples} & \textbf{Original $\kappa$} & \textbf{Final $\kappa$} & \textbf{Improvement} & \textbf{\# Human Samples} & \textbf{\# Fixes} \\
    \midrule
    FQ   & 5   & 0.55 & 0.75 & +0.20 & 5  & 0 \\
    PQ   & 11  & 0.50 & 0.56 & +0.06 & 8  & 0 \\
    RQ   & 331 & 0.63 & 0.68 & +0.05 & 25 & 20 \\
    OS   & 10  & -0.02 & -0.00 & +0.01 & 1  & 0 \\
    SS   & 119 & 0.79 & 0.80 & +0.01 & 6  & 3 \\
    RT   & 1   & -0.00 & 0.00 & +0.00 & 0  & 0 \\
    LO   & 51  & 0.80 & 0.80 & +0.00 & 1  & 0 \\
    RR & 4   & -0.00 & -0.00 & +0.00 & 0  & 0 \\
    SI  & 2   & 0.67 & 0.67 & +0.00 & 1  & 0 \\
    SR  & 1   & 0.00 & 0.00 & +0.00 & 0  & 0 \\
    \bottomrule
  \end{tabular}}
\end{table}

\section{Discussion}
\subsection{The LLMs are not always an expert for deductive coding.}

Through a systematic comparison of LLMs with varying parameter sizes and a series of BERT-based models across different datasets and codebooks, we found that LLMs do not adapt well to deductive coding tasks. On high-frequency codes, their performance was less consistent than that of fine-tuned small transformers, and on rare yet theory-critical codes, inter-rater agreement often dropped to nearly zero. These results are broadly aligned with \citeauthor{huang2025s}'s findings \cite{huang2025s}, and partially consistent with the results of \citet{long2024evaluating}, though they hold the view that LLMs can adapt well in deductive coding. Results of \citet{liu2025qualitative} in coding practice assignment submissions were also consistent with our findings. However, our findings diverge from parts of \citeauthor{liu2025qualitative}’s results \cite{liu2025qualitative}, which reported per-code kappa values in coding student observations and showed relatively strong performance even on rare codes within the context of scientific observations made by students. Prior studies did not provide an in-depth analysis of why LLMs underperformed on these codes. For example, \citet{na2025llm} reported only overall kappa values without offering per-code reliability. Our study systematically compared LLMs and small transformer-based models and demonstrated the core limitation of LLMs. This limitation is particularly consequential for learning analytics, where several rare codes may constitute the primary source of important theoretical insights \cite{bistolfi2025towards}. As emphasized in the literature, the value of qualitative coding lies in its ability to uncover processes and mechanisms in human learning that remain hidden in quantitative statistics \cite{papamitsiou2014learning}. Our findings show that current LLMs undermine precisely this aim: by systematically neglecting rare but theoretically critical codes. Consistent with broader text-classification evidence \cite{edwards2024language}, we observed that fine-tuned BERT was more reliable for deductive coding tasks than LLMs, which echoes the point of \citet{ashwin2023using}. Entering the era of GenAI does not necessarily mean that LLMs always function as expected. Beyond this technical uncertainty, issues of feasibility, sustainability, and cost-effectiveness remain central obstacles to the practical application of learning analytics in real educational contexts \cite{tsai2018sheila}. \textbf{Therefore, our findings also enable researchers in the field of learning analytics to delineate more precisely and clearly the capability boundaries between smaller models, such as BERT, and larger-scale models, and to encourage careful reflection on their appropriate use and trade-offs. }

After establishing that LLMs are ill-suited for deductive coding tasks, we use empirical data to explore the reasons behind their poor performance. To our knowledge, this issue has not been adequately discussed in prior studies. Our results further resonate with \citeauthor{reif2024beyond}’s findings \cite{reif2024beyond} on label bias in LLM-based classification tasks, and we also observed that LLMs did not consistently adhere to codebook definitions but instead relied on their own interpretation of code names, echoing the observations of \citet{halterman2024codebook}. We contend that this limitation may be partly rooted in the architecture of GenAI: decoder-only LLMs are naturally optimized for open-ended generation \cite{benayas2025comparative}, whereas encoder-only/encoder–decoder architectures such as BERT or T5 are inherently better suited for constrained  \cite{devlin2019bert,raffel2020exploring}, classification-like tasks such as deductive coding. Moreover, our analysis revealed systematic Type II errors similar to those reported by \citet{song2024large}, where LLMs frequently under-detected or rejected true positive instances. Taken together, our findings suggest that LLMs should not be treated as primary annotators for deductive, codebook-guided coding — particularly when code definitions are semantically overlapping. Instead, they are better positioned as assistants: triaging difficult cases, generating rationales for human adjudication, and supporting codebook auditing, rather than replacing the interpretive role of human coders.

\subsection{'Human in the Loop' is necessary in deductive coding}

Our results demonstrate that human expert adjudication substantially improves the reliability of deductive coding: while MacBERT and LLM-only predictions plateaued at $\kappa=0.62$, expert adjudication increased overall reliability to $\kappa=0.66$, with the largest gains observed on the rare but theoretically critical code \textbf{FQ} (+0.20).
Human–AI collaboration has proven effective in inductive or thematic analysis by externalizing under-specified criteria and iteratively aligning models with researcher intent \cite{gebreegziabher2023patat,gao2024collabcoder,oksanen2025llmcode,xu2025tama,dai2023llm,bryda2024words}). We found it to be applicable in deductive coding tasks, particularly those in which the theoretical insight and interpretive depth of human experts remain indispensable. 
Qualitative coding is fundamentally a theory-driven interpretive process \cite{fife2024deductive}. Prior work stresses that researchers must retain ultimate authority \cite{zhang2023qualigpt} and that interpretive depth cannot be automated \cite{sabbaghan2024exploring}, yet these claims have rarely been substantiated with empirical evidence. Our results provide such evidence, showing that even limited adjudication by human experts safeguards validity, especially on rare but theoretically critical codes. In our workflow, frequent codes can be delegated to smaller transformer-based models with acceptable reliability, whereas low-confidence predictions and rare codes are escalated for expert adjudication. This design resonates with the logic of selective prediction in active learning \cite{settles2009active}, but situates it within the epistemic requirements of qualitative analysis. 

Rather than a finalized solution, our workflow should be viewed as a proof of concept that illustrates possible directions for designing human–AI collaborative coding. In this design, role allocation is explicit: small transformer-based models handle frequent codes with acceptable reliability, LLMs provide rationales and triage support, and human experts adjudicate cases that are either low-confidence or theoretically critical. The routing mechanism combines a confidence threshold with a sparsity condition, ensuring that rare or ambiguous cases are systematically escalated. 

From this perspective, the “loop” in human-in-the-loop should be understood not as a simple corrective step at the end of model predictions, but as a designed workflow that structures the division of labour between humans and AI. Finally, the loop also feeds back into theory: human adjudication revealed semantically overlapping codes that confused both models and coders, highlighting the need for iterative codebook refinement \cite{macqueen1998codebook}. Taken together, these elements indicate that human–AI collaboration in deductive coding is best understood not as model substitution, but as workflow orchestration that embeds human judgment at its core. \textbf{For the learning analytics community, such workflows not only make large-scale coding more feasible by automating routine categories, but also safeguard the theoretical integrity of the rare codes. In this way, human-in-the-loop design benefits both efficiency of everyday coding practice and credibility of insights drawn for learning analytics stakeholders.}

\section{Limitations and Future Work}
This study has several limitations. First, our datasets and code schemes focused on a specific educational domain and language, limiting external validity across domains and languages. Second, the LLMs pool and transformer baselines are limited; we did not exhaust domain-adapted, or tool-augmented variants, nor perform extensive fine-tuning on LLMs. Third, we relied on standard hyperparameters and a single imbalance-mitigation recipe with limited ablations. We did not systematically tune learning rate, batch size, weight decay or gradient clipping strategies. 
Nor did we compare alternative tail-focused objectives (e.g., focal, LDAM, logit-adjusted losses), calibration and per-code thresholding or domain-adaptive pretraining. Accordingly, our transformer results should be read as points on a broad design surface rather than fully optimised ceilings. Future studies could broaden coverage across domains, genres, and languages, and move beyond a single imbalance recipe by systematically optimising transformer training and comparing tail-oriented objectives such as focal, class-balanced, logit-adjusted, or AUC-max losses, together with calibration and per-code decision thresholds. Evaluation may be extended with prevalence- or bias–adjusted reliability, PR AUC, and utility-weighted scores that emphasise theory-critical tails. In terms of workflow design, future coding practice should thus aim for a synergistic 1+1>2 collaboration, where human interpretation and AI support jointly enhance coding reliability, theoretical insight, and data understanding.

\begin{acks}
We thank the anonymous reviewers and the meta-reviewer for their constructive and detailed feedback, which helped us strengthen the framing, clarify key arguments, and improve the presentation of results. We also thank all co-authors for their substantial contributions throughout the study and for their careful revisions that improved this manuscript.
\end{acks}

\bibliographystyle{ACM-Reference-Format}
\bibliography{main_draft}
\settopmatter{printacmref=true} 
\appendix

\section*{Appendix}

\section{Full Codebook}
\label{appendix:codebook_history}

\begin{table*}[htbp]
  \centering
  \caption{Full coding scheme for history taking.}
  \label{tab:codebook_history_full}
  \resizebox{\textwidth}{!}{
  \begin{tabular}{p{3cm}p{9cm}p{6cm}}
    \toprule
    \textbf{Code} & \textbf{Definition} & \textbf{Example} \\
    \midrule
    Pathophysiologic Question (PQ) & Hypothesis-driven questions grounded in pathophysiologic reasoning (etiology, triggers, family history, etc.). & ``Does the pain get worse when climbing stairs?'' \\
    Relevant Response (RR) & Recognition and follow-up on diagnostically significant patient information. & ``For how long have you been taking contraceptives?'' \\
    Summarizing \& Integrating (SI) & Synthesizing and organizing patient information in a coherent, structured way. & ``You had sudden right chest pain this morning after blowing a balloon...'' \\
    Logical Organization (LO) & Questions that reflect logical exploration of associated symptoms and severity. & ``Do you have shortness of breath or cyanosis?'' \\
    Specifying Symptoms (SS) & Systematic exploration of symptom characteristics (onset, duration, location, etc.). & ``What triggered your chest pain today?'' \\
    Routine Question (RQ) & Standard questions (demographics, past medical history, ROS, etc.). & ``Do you have any other discomfort?'' \\
    Summarizing \& Restating (SR) & Restating collected information without reorganization. & ``You came to the hospital for chest pain and shortness of breath...'' \\
    Checking (CK) & Confirming or clarifying patient-reported information. & ``So the pain lasted about 2 minutes, right?'' \\
    Repeating Question (RT) & Asking about the same information multiple times (often due to limited knowledge). & ``Do you have shortness of breath?'' (after already asking similarly) \\
    Fuzzy Question (FQ) & Vague, open-ended prompts repeated within the same history domain. & ``Any other diseases?'' \\
    Chitchat (CC) & Social talk, reassurance, greetings, or transitional explanations. & ``Hello, I’m Dr. Zhang. Don’t worry, take your time.'' \\
    Off-topic Statement (OS) & Irrelevant, illogical, or incomplete statements. & ``Did the balloon get sucked in?'' \\
    \bottomrule
  \end{tabular}}
\end{table*}

\begin{table*}[htbp]
  \centering
  \small
  \caption{Codebook 2: Integrated question types and mechanisms (adapted from \citeauthor{cheng2025asking}.}
  \label{appendix:codebook_combined}
  \resizebox{\textwidth}{!}{
  \begin{tabular}{lll}
    \toprule
    \textbf{Code} & \textbf{Definition} & \textbf{Example} \\
    \midrule
    \multicolumn{3}{c}{\textbf{Question Types}} \\
    \midrule
    Verification & Confirm a fact or event & ``You mean the first paragraph still need to expand?'' \\
    Disjunctive & Choose among a set of options & ``Should I first do x, or do y?'' \\
    Concept Completion & Identify/complete missing element & ``What is the correct spelling of vechicle [sic]?'' \\
    Example & Request an instance exemplifying a category & ``Can you give examples on how to use differentiation?'' \\
    Feature Specification & Ask about qualitative attributes of an entity & ``Can you explain applications of scaffolding to me?'' \\
    Definition & Clarify the meaning of a concept or term & ``What is the definition of differentiation?'' \\
    Comparison & Explore similarities/differences between entities & ``What's the difference between scaffolding and instructional approaches?'' \\
    Causal Consequence & Ask about the effects of an event or state & ``What are the potential consequences of overusing AI in education?'' \\
    Instrumental & Ask about the means to accomplish a goal & ``How to apply AI in education?'' \\
    Enablement & Ask about enabling resources or conditions & ``Can you give me some ideas to support differentiation?'' \\
    Judgmental & Evaluate an idea or seek advice & ``How would you rate the quality of this essay?'' \\
    Assertion & Indicate lack of knowledge or understanding & ``uh, I don't know what revision is needed...'' \\
    Indirect Request & Polite form of asking for action & ``Could you please give me feedback on my essay?'' \\
    Direct Request & Commanding/direct form of asking for action & ``Give me the feedback of my essay.'' \\
    \midrule
    \multicolumn{3}{c}{\textbf{Question Mechanisms}} \\
    \midrule
    Knowledge Deficit & Asked when knowledge is incomplete, missing, or contradictory & ``How does AI enhance scaffolding?'' \\
    Social Coordination & Questions to coordinate actions among participants & ``Please help me to make this clearer.'' \\
    Common Ground & Ensure shared understanding or confirm beliefs & ``Do you think I wrote a lot at the beginning?'' \\
    Conversation Control & Manage the flow of conversation (greetings, rhetorical, etc.) & ``Good morning! Can you give me advice about my article?'' \\
    \bottomrule
  \end{tabular}}
\end{table*}

\section{Detailed Table and Figure of Results}
\label{appendix:bert_optimized}
\begin{table}[htbp]
  \centering
  \caption{Performance of optimized transformer models on $\kappa$.}
  \label{tab:bert_optimized}
  \begin{tabular}{lccc}
    \toprule
    \textbf{Code} & \textbf{BERT} & \textbf{RoBERTa-wwm-ext} & \textbf{MacBERT} \\
    \midrule
    SS   & 0.884 & 0.874 & 0.881 \\
    LO   & 0.833 & 0.840 & 0.842 \\
    RQ   & 0.876 & 0.872 & 0.871 \\
    PQ   & 0.563 & 0.623 & 0.586 \\
    RR & 0.555 & 0.565 & 0.604 \\
    SI  & 0.843 & 0.863 & 0.832 \\
    SR  & 0.460 & 0.363 & 0.460 \\
    CK   & 0.382 & 0.387 & 0.392 \\
    FQ   & 0.469 & 0.567 & 0.544 \\
    RT   & 0.265 & 0.208 & 0.248 \\
    CC   & 0.918 & 0.916 & 0.916 \\
    OS   & 0.184 & 0.239 & 0.197 \\
    \bottomrule
  \end{tabular}
 \caption*{\footnotesize Note: Models were trained with \textit{loss reweighting} and \textit{weighted sampling} to mitigate class imbalance.}
\end{table}

\begin{table}[htbp]
  \centering
  \caption{Experiment 1: Cohen’s $\kappa$ for LLMs under reduced coding scope.}
  \label{tab:exp1_llm}
  \begin{tabular}{lcc}
    \toprule
    \textbf{Code} & \textbf{GPT-4o} & \textbf{GPT-4o-mini} \\
    \midrule
    CC   & 0.71 & 0.69 \\
    CK   & 0.38 & 0.39 \\
    RT   & 0.16 & 0.00 \\
    FQ   & 0.56 & 0.50 \\
    OS   & 0.31 & 0.26 \\
    PQ   & 0.55 & 0.50 \\
    RR & 0.39 & 0.19 \\
    SI  & 0.29 & 0.62 \\
    SR  & 0.32 & 0.33 \\
    \bottomrule
  \end{tabular}
\end{table}

\begin{table}[htbp]
  \centering
  \caption{Experiment 2: Cohen’s $\kappa$ for binary codewise judgments.}
  \label{tab:exp2_llm}
  \begin{tabular}{lcc}
    \toprule
    \textbf{Code} & \textbf{GPT-4o} & \textbf{GPT-4o-mini} \\
    \midrule
    PQ   & 0.41 & 0.42 \\
    RR & 0.08 & 0.09 \\
    SI  & 0.49 & 0.51 \\
    SR  & 0.11 & 0.09 \\
    CK   & 0.15 & 0.15 \\
    FQ   & 0.37 & 0.35 \\
    RT   & 0.04 & 0.06 \\
    OS   & 0.13 & 0.14 \\
    RQ   & 0.00 & 0.00 \\
    SS   & 0.00 & 0.00 \\
    LO   & 0.00 & 0.00 \\
    CC   & 0.54 & 0.57 \\
    \bottomrule
  \end{tabular}
\caption*{\footnotesize Note: Each sample required 12 binary (yes/no) decisions. }

\end{table}

\begin{figure*}[htbp]
  \centering
  \begin{subfigure}{0.45\linewidth}
    \includegraphics[width=\linewidth]{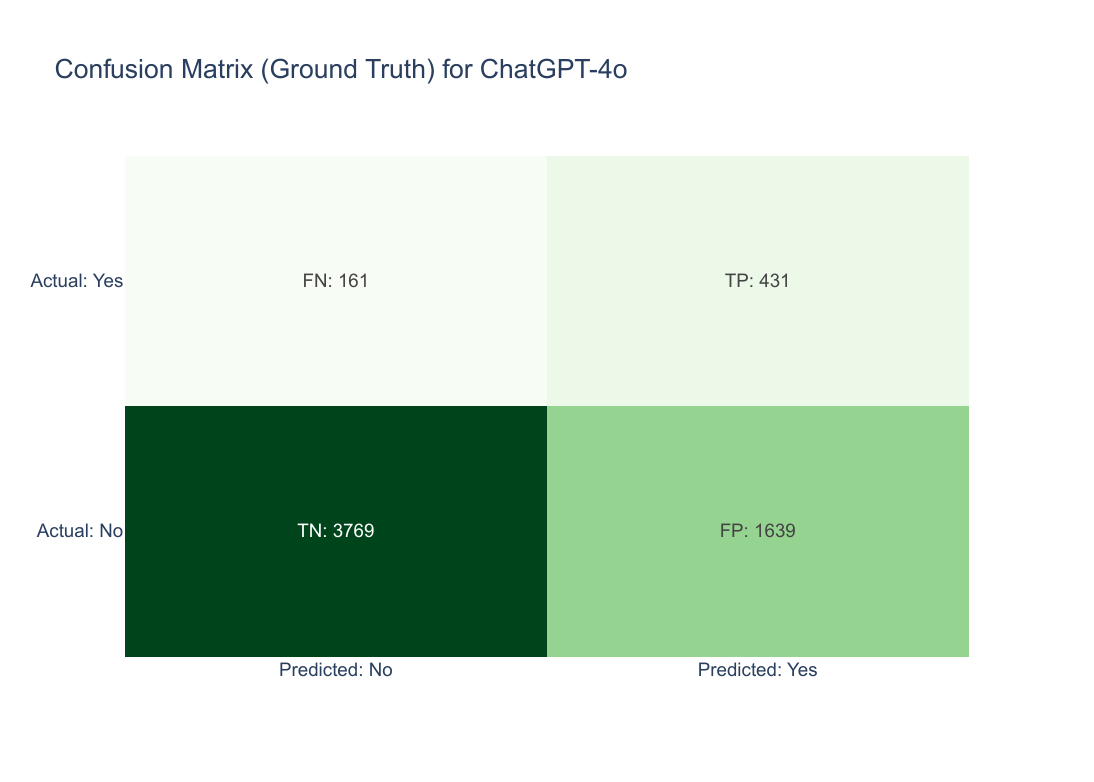}
    \caption{GPT-4o}
    \label{fig:exp2_confusion_4o}
  \end{subfigure}
  \hfill
  \begin{subfigure}{0.45\linewidth}
    \includegraphics[width=\linewidth]{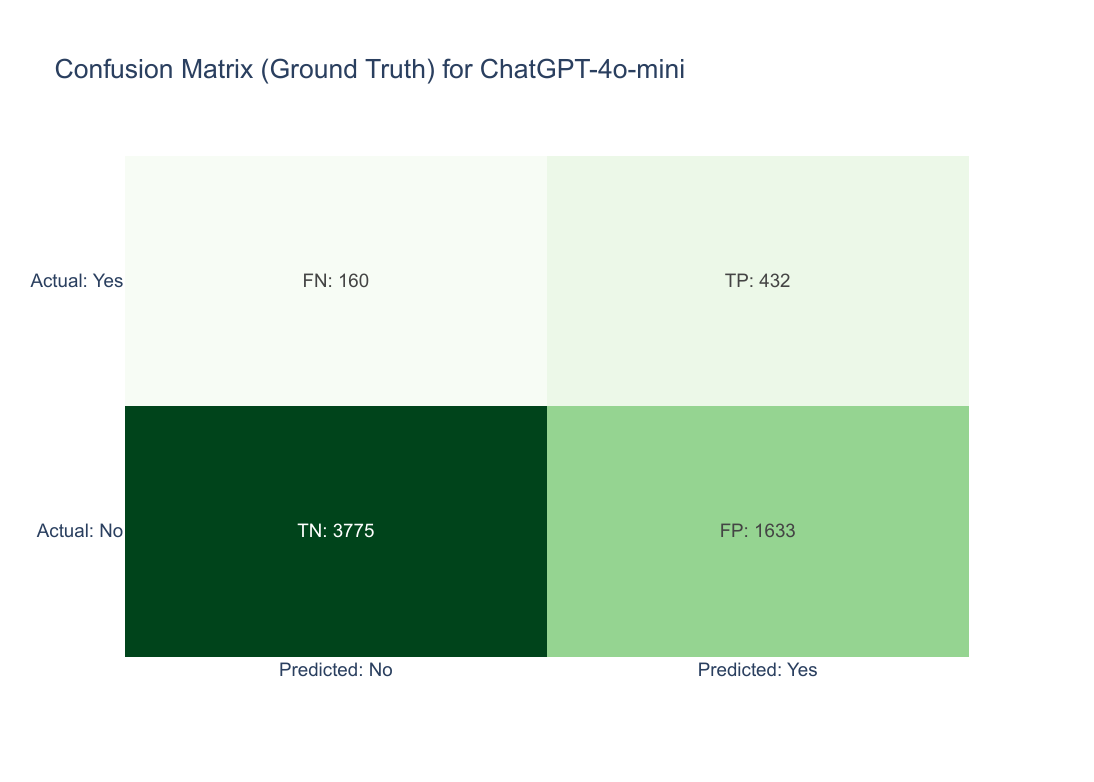}
    \caption{GPT-4o-mini}
    \label{fig:exp2_confusion_4o_mini}
  \end{subfigure}
  \caption{Error distribution in Experiment~2 (binary judgments). Both models exhibit a strong bias toward negative predictions.}
  \label{fig:exp2_confusion}
\end{figure*}

\begin{figure*}[htbp]
  \centering
  \begin{subfigure}{0.45\linewidth}
    \includegraphics[width=\linewidth]{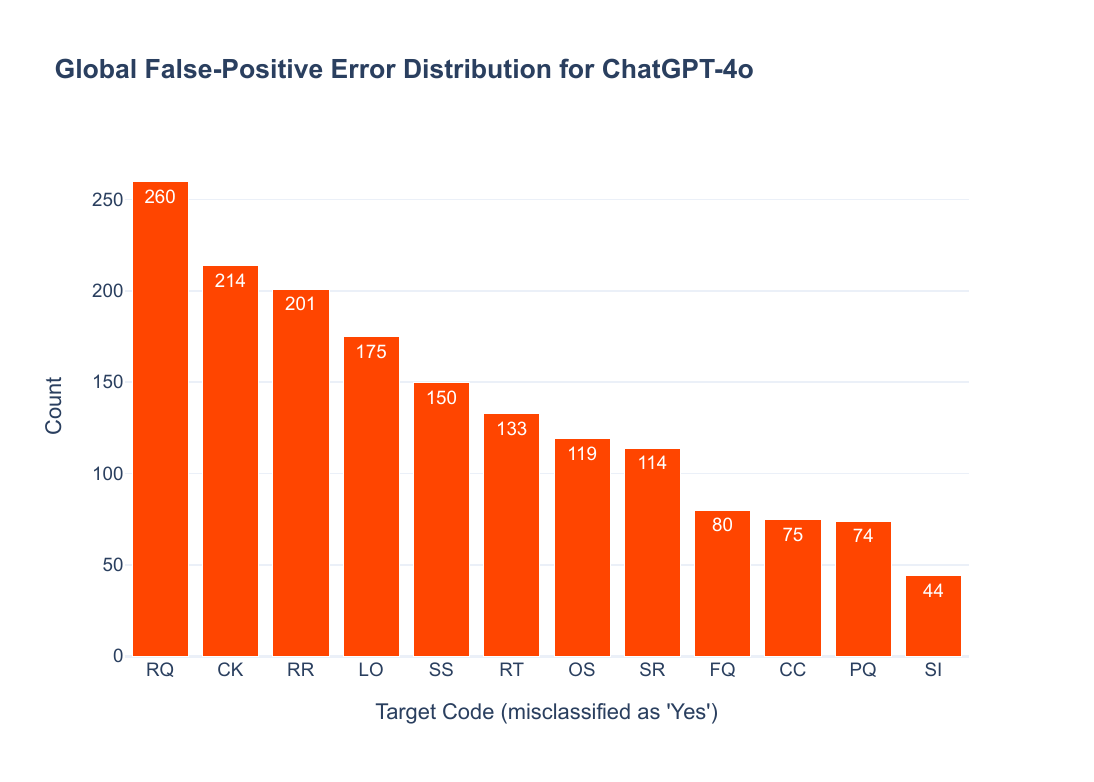}
    \caption{GPT-4o}
    \label{fig:exp2_error_distribution_4o}
  \end{subfigure}
  \hfill
  \begin{subfigure}{0.45\linewidth}
    \includegraphics[width=\linewidth]{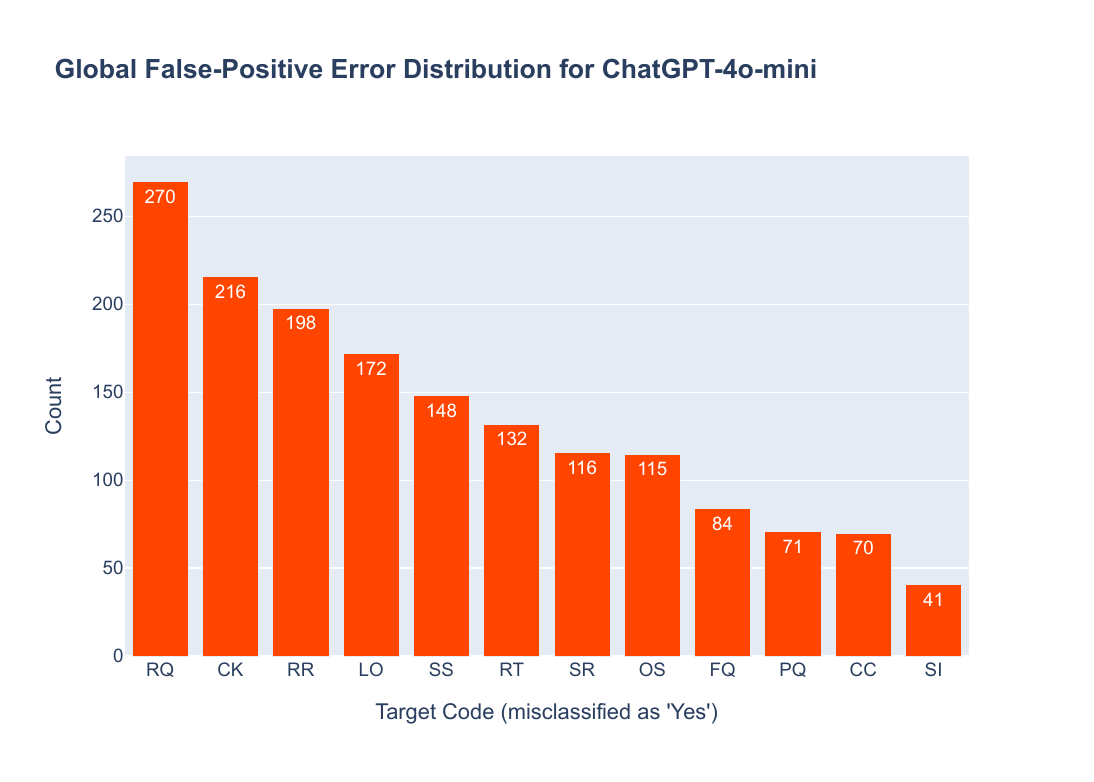}
    \caption{GPT-4o-mini}
    \label{fig:exp2_error_distribution_4o_mini}
  \end{subfigure}
  \caption{False Positive Error distribution in Experiment~2 (binary judgments).}
  \label{fig:exp2_fp_error_distribution_in_code}
\end{figure*}

\begin{table*}[htbp]
  \centering
  \caption{LLM performance (Cohen’s $\kappa$) on \textbf{Question Types} codebook. Support indicates the number of annotated instances per code.}
  \label{tab:supp_question_types}
  \resizebox{0.9\textwidth}{!}{
  \begin{tabular}{lccccc r}
    \toprule
    \textbf{Code} & \textbf{Qwen3-8B} & \textbf{Qwen3-30B} & \textbf{DeepSeek-V3.1} & \textbf{GPT-4o-mini} & \textbf{GPT-4o} & \textbf{Support} \\
    \midrule
    Judgmental         & 0.3169 & 0.2728 & 0.4441 & 0.2618 & 0.3469 & 163 \\
    Direct Request     & 0.1207 & 0.5225 & 0.6038 & 0.6326 & 0.6485 & 161 \\
    Instrumental       & 0.3624 & 0.5814 & 0.8229 & 0.7001 & 0.6695 & 106 \\
    Indirect Request   & 0.2077 & 0.4909 & 0.4909 & 0.5068 & 0.6349 & 61 \\
    Definition         & 0.5824 & 0.5763 & 0.7868 & 0.4857 & 0.6543 & 34 \\
    Feature Specification & 0.1822 & 0.2626 & 0.4858 & 0.1492 & 0.2021 & 33 \\
    Assertion          & 0.2129 & 0.2951 & 0.3225 & 0.2409 & 0.2912 & 23 \\
    Concept Completion & 0.1743 & 0.1116 & 0.6489 & 0.0898 & 0.2326 & 21 \\
    Off Topic          & 0.0000 & 0.0000 & 0.0000 & 0.0000 & 0.0000 & 14 \\
    Causal Consequence & 0.4111 & 0.5925 & 0.7215 & 0.4969 & 0.5150 & 12 \\
    Enablement         & 0.1416 & 0.3273 & 0.3223 & 0.3432 & 0.3932 & 11 \\
    Verification       & 0.0919 & 0.0800 & 0.2814 & 0.2137 & 0.3110 & 11 \\
    Example            & 0.7741 & 0.7012 & 0.9324 & 0.5514 & 0.4400 & 7 \\
    Comparison         & 0.2738 & 0.5196 & 0.4927 & 0.5196 & 0.5196 & 5 \\
    Disjunctive        & 0.3299 & 0.3299 & 0.6638 & 0.5410 & 0.6113 & 4 \\
    Expectational      & 0.0000 & 0.0000 & 0.0000 & 0.0000 & 0.0000 & 2 \\
    \bottomrule
  \end{tabular}}
\end{table*}

\begin{table}[htbp]
  \centering
  \caption{LLM performance (Cohen’s $\kappa$) on \textbf{Question Mechanisms} codebook.}
  \label{tab:supp_mechanisms}
  \resizebox{0.9\linewidth}{!}{
  \begin{tabular}{lccccc r}
    \toprule
    \textbf{Mechanism} & \textbf{Qwen3-8B} & \textbf{Qwen3-30B} & \textbf{DeepSeek-V3.1} & \textbf{GPT-4o-mini} & \textbf{GPT-4o} & \textbf{Support} \\
    \midrule
    Social Coordination  & 0.3904 & 0.5670 & 0.5863 & 0.3863 & 0.5334 & 284 \\
    Knowledge Deficit    & 0.5148 & 0.6647 & 0.7341 & 0.4766 & 0.6696 & 225 \\
    Common Ground        & 0.1949 & 0.4900 & 0.4942 & 0.3709 & 0.4180 & 55  \\
    Conversation Control & 0.0000 & 0.0000 & 0.0000 & 0.0000 & 0.0000 & 9   \\
    \bottomrule
  \end{tabular}}
\end{table}

\begin{figure}[htbp]
  \centering
  \includegraphics[width=0.7\linewidth]{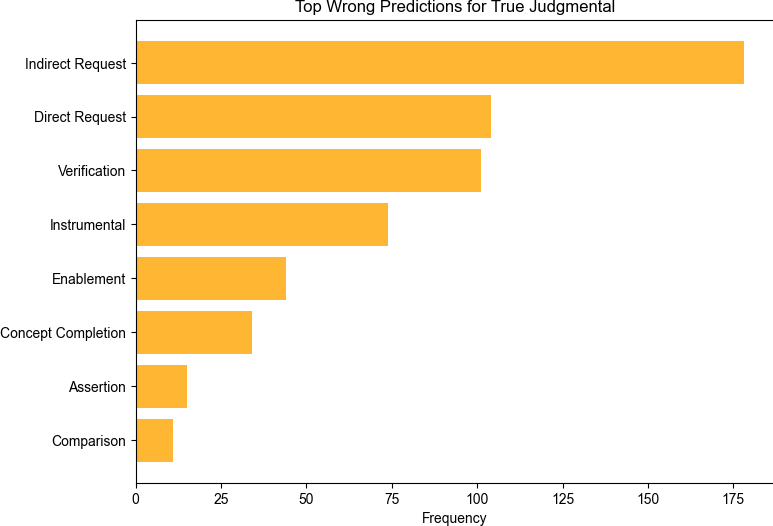}
  \caption{False Negatives Distribution in 'Judgemental'.}
  \Description{False Negatives Distribution in 'Judgemental'.}
  \label{fig:fn_judgemental}
\end{figure}

\begin{figure}[htbp]
  \centering
  \begin{subfigure}{0.45\linewidth}
    \includegraphics[width=\linewidth]{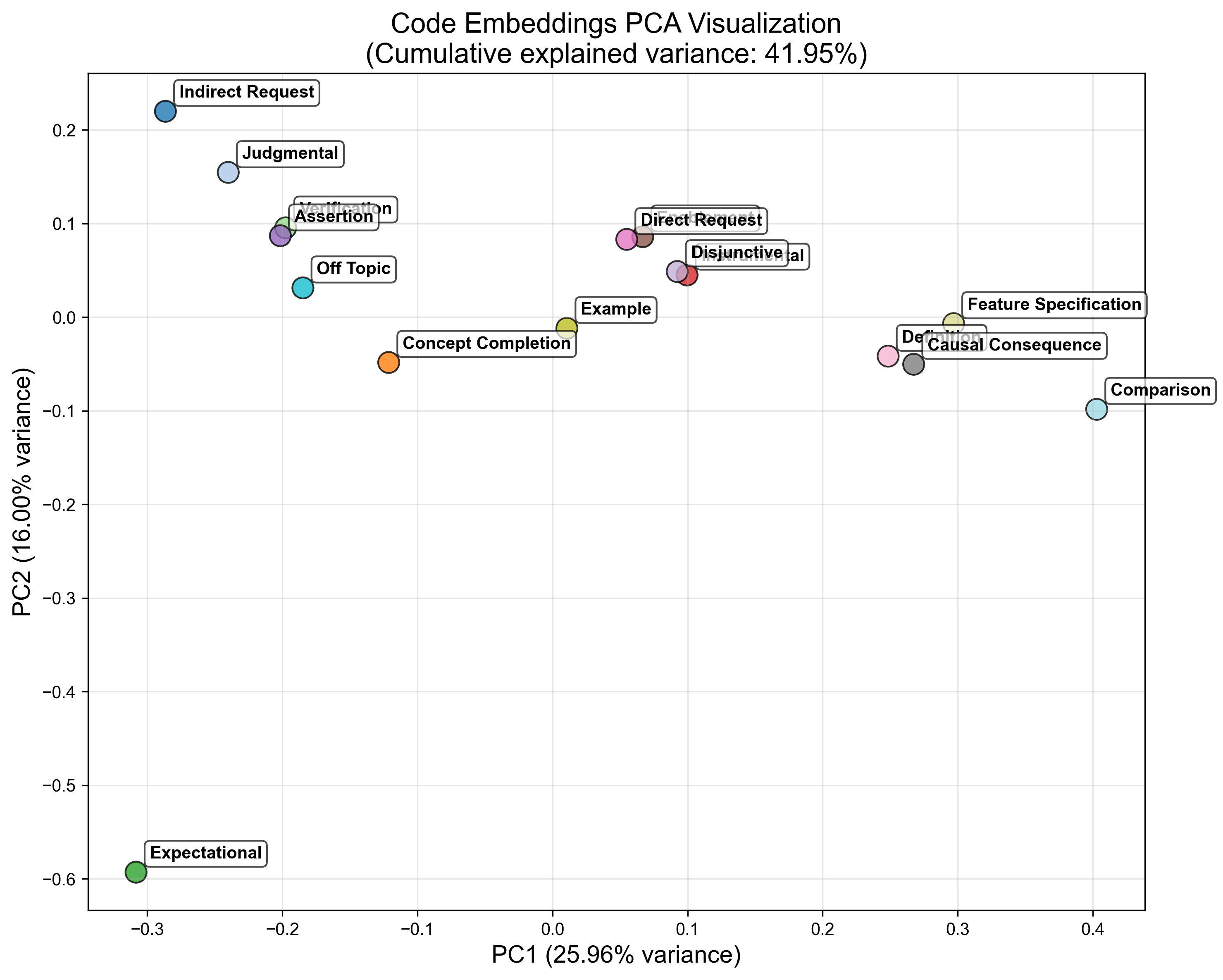}
    \caption{PCA Scatter of Codes in Question Type.}
    \label{fig:qt_pca_scatter}
  \end{subfigure}
  \hfill
  \begin{subfigure}{0.45\linewidth}
    \includegraphics[width=\linewidth]{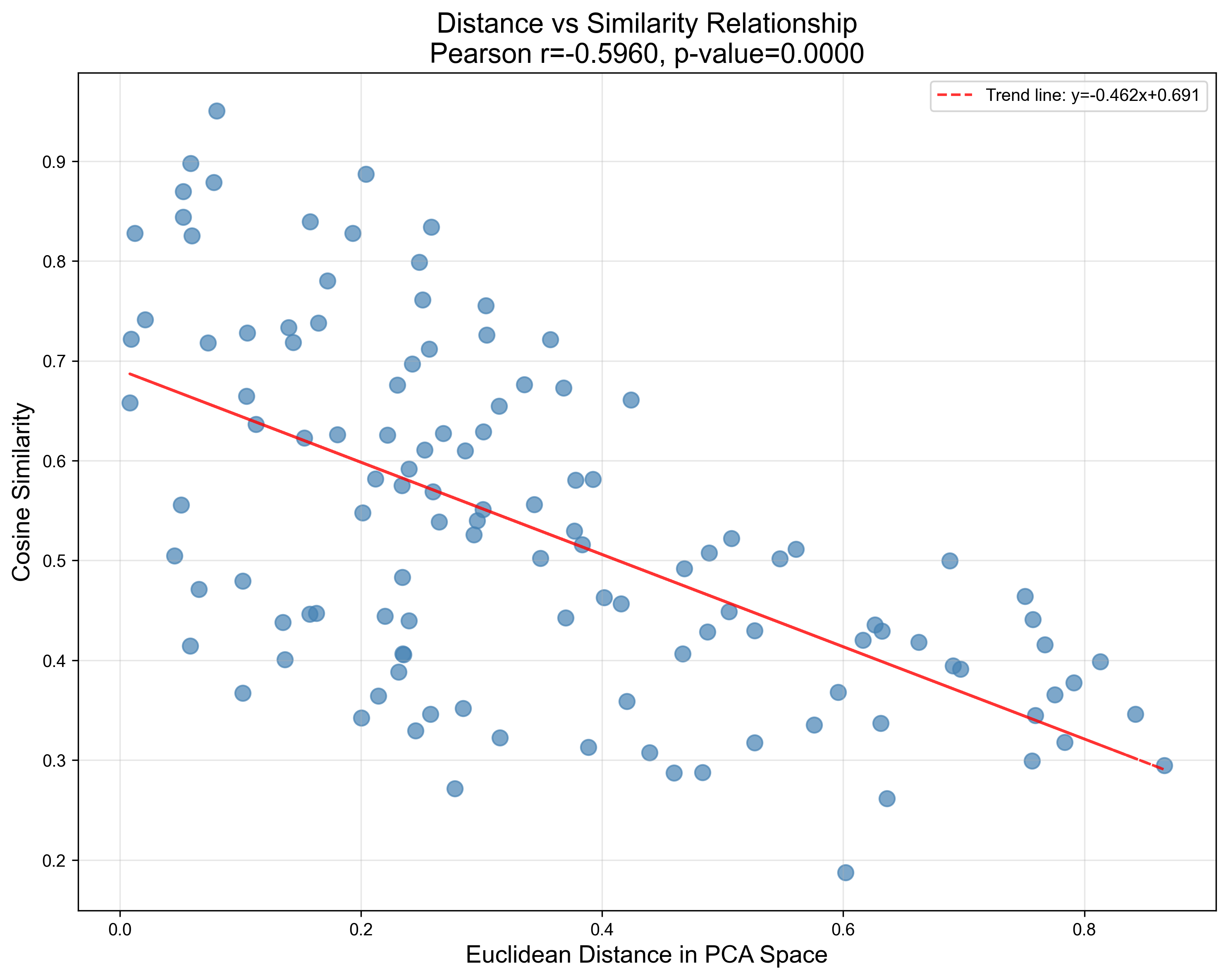}
    \caption{Similarity-Distance}
    \label{fig:qt_similarity_distance}
  \end{subfigure}
  \caption{Embedding Similarity Analysis Result in Question Type.}
  \label{fig:qt_embedding_similarity}
\end{figure}

\begin{figure}[htbp]
  \centering
  \begin{subfigure}{0.45\linewidth}
    \includegraphics[width=\linewidth]{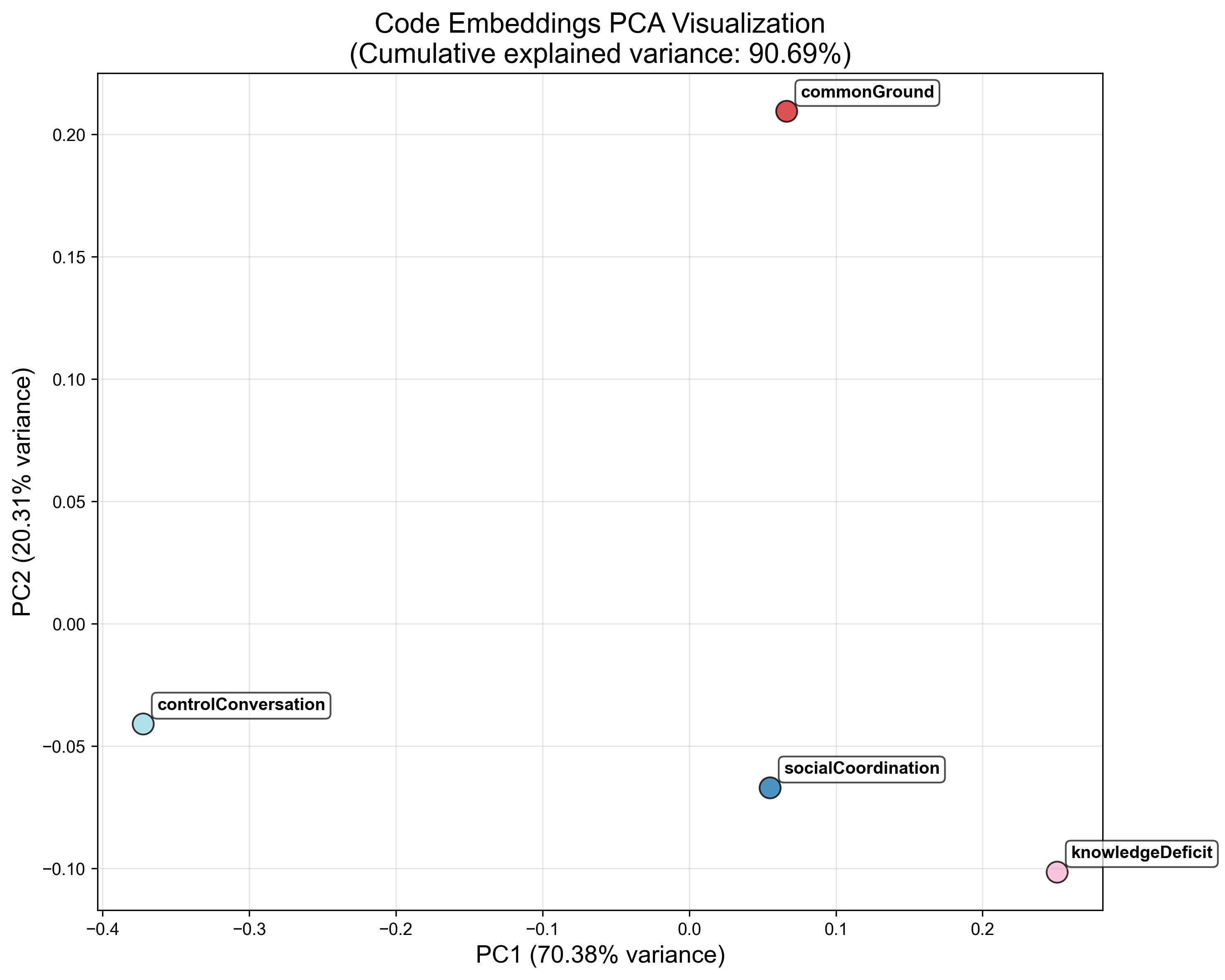}
    \caption{PCA Scatter of Codes in Mechanism.}
    \label{fig:mechanism_pca_scatter}
  \end{subfigure}
  \hfill
  \begin{subfigure}{0.45\linewidth}
    \includegraphics[width=\linewidth]{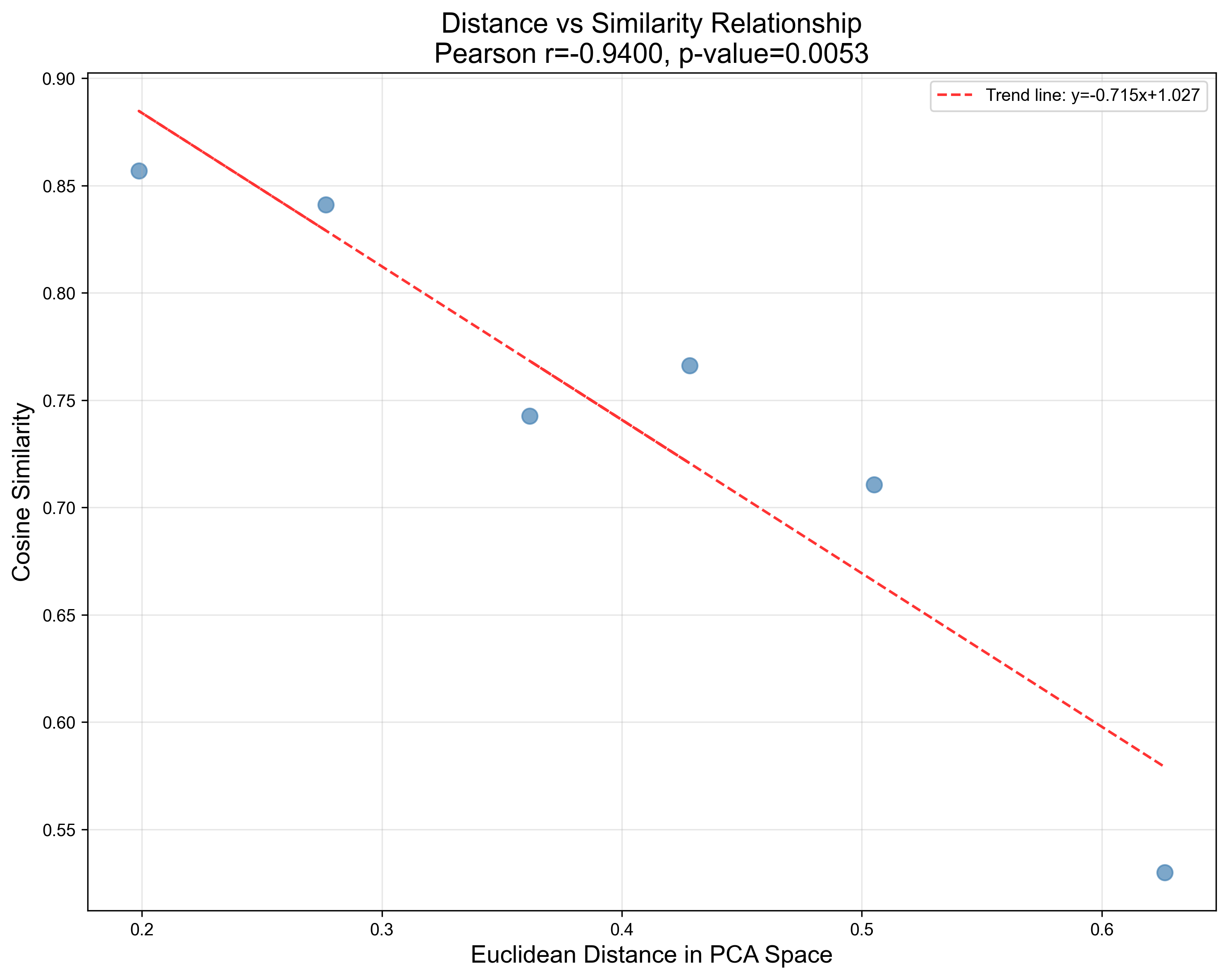}
    \caption{Similarity-Distance}
    \label{fig:mechanism_similarity_distance}
  \end{subfigure}
  \caption{Embedding Similarity Analysis Result in Mechanism.}
  \label{fig:mechanism_embedding_similarity}
\end{figure}

\section{Prompt Examples}
\label{sec:appendix_prompts}
This appendix provides the exact prompt templates used in our experiments. 
Each task type is illustrated with a representative prompt. 
Variables in \{\} are placeholders dynamically filled with data, e.g., dialogue context, student question, or code definitions.

\subsection{Prompt for Systematic Model Comparison (Medical Dialogue)}

\begin{tcolorbox}[colback=gray!5!white,colframe=gray!70!black,title={Prompt Example: Systematic Comparison}]
\small
\textbf{System Prompt} \\
You are an expert in medical dialogue coding. Your task is to classify student questions according to the given coding scheme.

\begin{itemize}
  \item \textbf{Coding Scheme:} \{codebook\}
  \item \textbf{Case Background:} \{case\_background\}
  \item \textbf{Coding Requirements:}
    \begin{itemize}
      \item Only analyze \textbf{student questions}.
      \item If multiple codes apply, list all codes separated by commas.
      \item Prefer the \textbf{single best-fitting code} whenever possible.
    \end{itemize}
  \item \textbf{Output Format:} Output only the code name(s), e.g., \texttt{RQ} or \texttt{RQ, CC}. Do \textbf{not} include explanations.
\end{itemize}

\medskip
\textbf{User Prompt} \\
Context: \{context\} \\
Student Question: \{question\} \\
Virtual Patient Response: \{answer\} \\

\textbf{Task:} Please code the Student Question.
\end{tcolorbox}

\subsection{Prompt for Binary Codewise Judgment}

\begin{tcolorbox}[colback=gray!5!white,colframe=gray!70!black,title={Prompt Example: Binary Judgment Task}]
\small
\textbf{System Prompt} \\
You are an expert in medical dialogue coding. Determine whether a student question belongs to a specific target code.

\begin{itemize}
  \item \textbf{Case Background:} \{case\_background\}
  \item \textbf{Target Code:} \{target\_code\}
  \item \textbf{Definition:} \{definition\}
  \item \textbf{Typical Examples:} \{examples\}
  \item \textbf{Key Features:} \{keywords\}
\end{itemize}

\textbf{Decision Rules:}
\begin{itemize}
  \item If the question matches the definition/features $\rightarrow$ output \texttt{Yes}.
  \item Otherwise $\rightarrow$ output \texttt{No}.
  \item Only answer with \texttt{Yes} or \texttt{No}.
\end{itemize}

\medskip
\textbf{User Prompt} \\
Context: \{context\} \\
Student Question: \{question\} \\
\textbf{Task:} Does this question belong to ``\{target\_code\}''? Answer \texttt{Yes} or \texttt{No}.
\end{tcolorbox}

\subsection{Prompt for Question Type Coding}

\begin{tcolorbox}[colback=gray!5!white,colframe=gray!70!black,title={Prompt Example: Question Type Coding}]
\small
Please analyse the following student utterance and classify it according to the \textbf{Question Types} coding scheme.

\begin{itemize}
  \item \textbf{Conversation Context:} \{context\}
  \item \textbf{Question Types Coding Scheme:} \{question\_types\_codebook\}
\end{itemize}

\textbf{Your Task:}  
Select the most appropriate Question Type from: \{question\_types\}.  

If two types are equally appropriate, output both separated by a slash (e.g., ``Verification/Instrumental'').  

\textbf{Output Format:} Respond with only the code name(s), e.g., ``Verification'' or ``Direct Request''.

\medskip
\textbf{Your Answer:} \_\_\_\_
\end{tcolorbox}

\subsection{Prompt for Question Mechanism Coding}

\begin{tcolorbox}[colback=gray!5!white,colframe=gray!70!black,title={Prompt Example: Question Mechanism Coding}]
\small
Please analyse the following student utterance and classify it according to the \textbf{Question-Generating Mechanisms} coding scheme.

\begin{itemize}
  \item \textbf{Conversation Context:} \{context\}
  \item \textbf{Mechanism Coding Scheme:} \{mechanisms\_codebook\}
\end{itemize}

\textbf{Your Task:}  
Identify the underlying mechanism that generated this question. Select from: \{mechanisms\}.  

If two mechanisms are equally appropriate, output both separated by a slash (e.g., ``Knowledge Deficit/Social Coordination'').  

\textbf{Output Format:} Respond with only the mechanism name(s), e.g., ``Knowledge Deficit'' or ``Social Coordination''.

\medskip
\textbf{Your Answer:} \_\_\_\_
\end{tcolorbox}

\section{Supplementary Error Analyses}
\label{appendix:impersonation}
\begin{figure*}[htbp]
  \centering
  \begin{subfigure}{0.4\linewidth}
    \includegraphics[width=\linewidth]{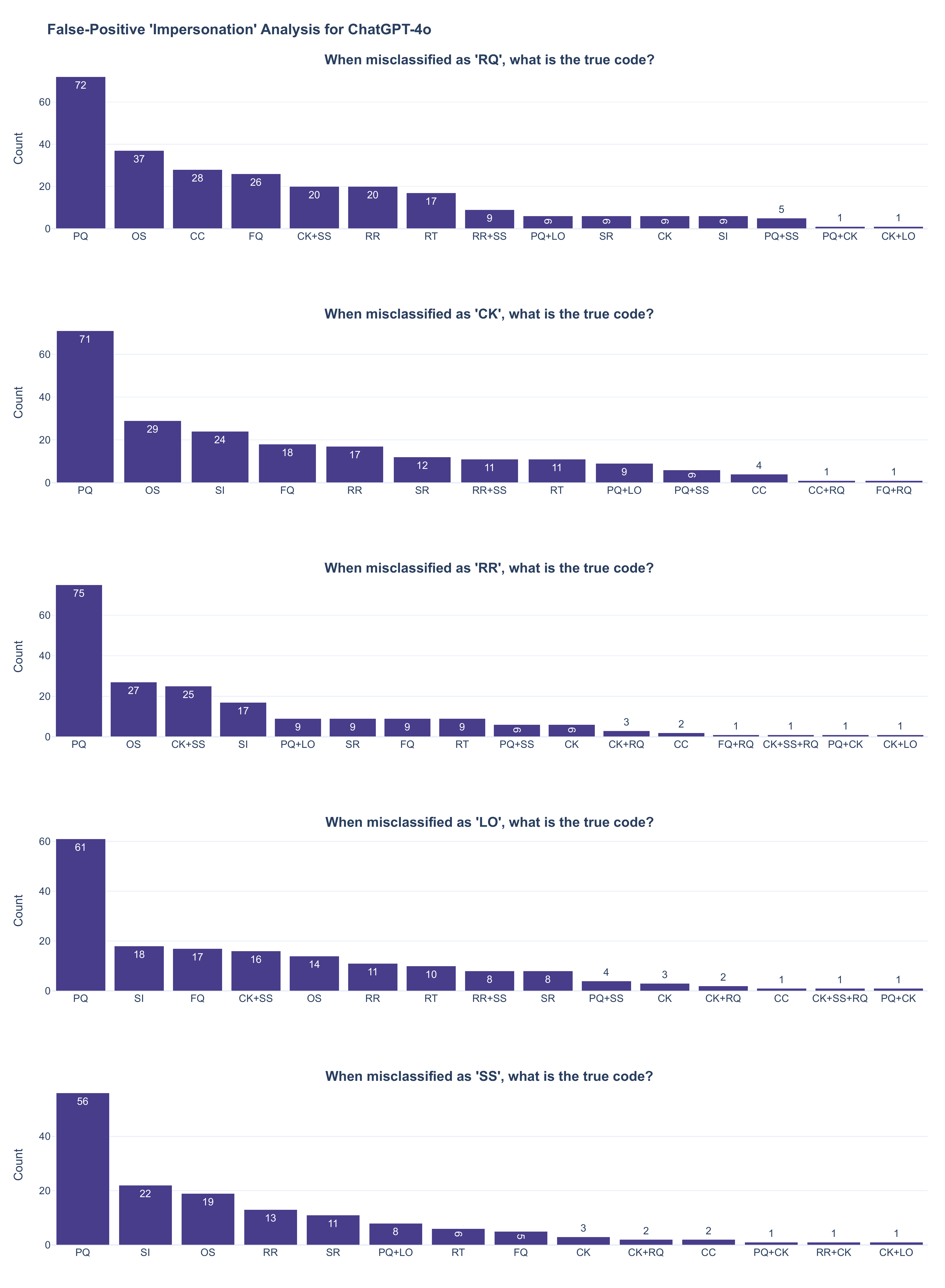}
    \caption{GPT-4o}
    \label{fig:exp2_impersonation_4o}
  \end{subfigure}
  \hfill
  \begin{subfigure}{0.4\linewidth}
    \includegraphics[width=\linewidth]{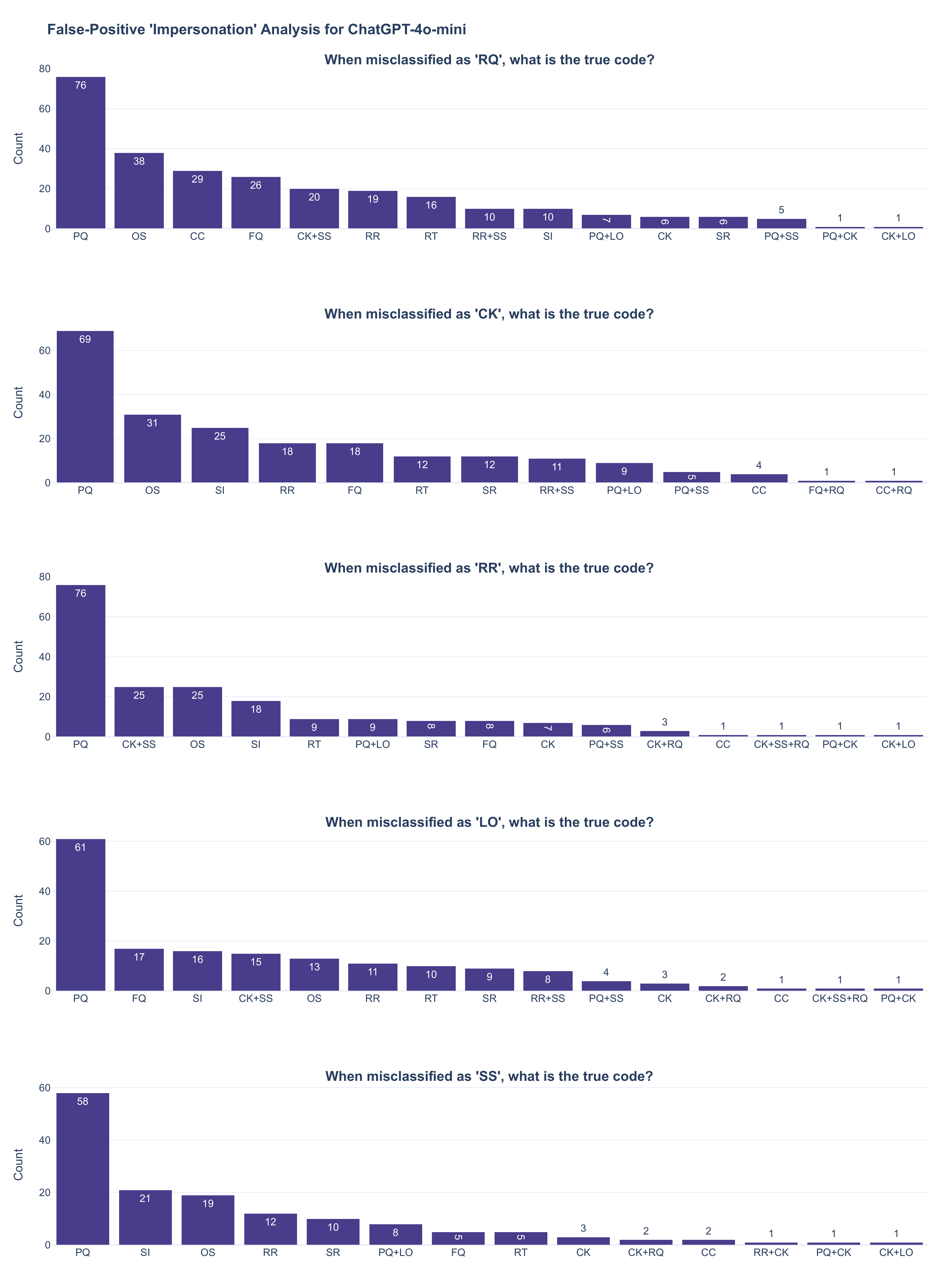}
    \caption{GPT-4o-mini}
    \label{fig:exp2_impersonation_4o_mini}
  \end{subfigure}
  \caption{False Positive Impersonationin Analysis Experiment~2 (binary judgments).}
  \label{fig:exp2_impersonation_in_code}
\end{figure*}

\end{document}